\documentclass[aps,pre,reprint,longbibliography,floatfix]{revtex4-2}
\usepackage[utf8]{inputenc}
\usepackage{amssymb,amsmath}
\usepackage{epsfig}
\usepackage{bm}
\usepackage{soul}
\usepackage{color}
\usepackage[mathlines]{lineno}% 
\usepackage{tabularx}
\usepackage{graphicx}% Include figure files

\usepackage{siunitx}

\usepackage{dcolumn}% Align table columns on decimal point

\newcommand\beq{\begin{equation}}
\newcommand\eeq{\end{equation}}
\newcommand\beqa{\begin{eqnarray}}
\newcommand\eeqa{\end{eqnarray}}
\newcommand{\dd}{\text{d}}

\newcommand{\al}{\alpha}
\newcommand{\rb}{\mathbf{r}}
\newcommand{\vb}{\mathbf{v}}

\begin{document}

\preprint{APS/123-QED}

\title{Tracer Diffusion in Granular Suspensions: Testing the Enskog Kinetic Theory with DSMC and Molecular Dynamics}% Force line breaks with \\

\author{Antonio M. Puertas}
 \email{apuertas@ual.es}
 \affiliation{Departamento de Química y Física, Universidad de Almer\'ia, 04120 Almer\'ia, Spain}
\author{Rub\'en G\'omez Gonz\'alez}%
 \email{ruben@unex.es}
\affiliation{%
 Departamento de
Did\'actica de las Ciencias Experimentales y las Matem\'aticas, Universidad de Extremadura, 10003 C\'aceres, Spain
}%

\date{\today}

\begin{abstract}
We investigate the diffusion of an intruder in a granular gas, with both components modeled as smooth hard spheres, both immersed in a low viscosity carrier fluid to form a particle-laden suspension. In this system, dissipative particle collisions coexist with the action of a solvent. The latter is modeled via a viscous drag force and a stochastic Langevin-like force proportional to the background fluid temperature. Building on previous kinetic theory and random-walk results of the tracer diffusion coefficient [R. Gómez González, E. Abad, S. Bravo Yuste, and V. Garzó, Phys. Rev. E \textbf{108}, 024903 (2023)], where random-walk predictions were compared with Chapman--Enskog results up to the second Sonine approximation, we assess the robustness of the Enskog framework by incorporating molecular dynamics (MD) simulations, using direct simulation Monte Carlo (DSMC) results as an intermediate reference. In particular, we focus on the intruder velocity autocorrelation function, considering intruders different masses (from 0.01 to 100 times the mass of the granular particles), and analyse the behavior of the intruder temperature and diffusion coefficient. Our results clarify the influence of the friction parameter and the conditions under which Enskog kinetic theory reliably describes intruder diffusion in granular suspensions.
\end{abstract}

\keywords{Enskog kinetic theory, intruder diffusion coefficient, granular suspensions, DSMC, molecular dynamics simulations}%Use showkeys class option if keyword
\maketitle

%\tableofcontents
\section{Introduction}

Granular materials are rarely found in isolation in nature; on the contrary, they are usually immersed in an interstitial fluid, such as air or water, forming suspensions \cite{G19}. In these systems, the fluid modifies the short-range interactions between the grains by introducing additional effects such as lubrication, resistance, and, occasionally, cohesion. This coupling between solid particles and fluid flow gives rise to complex rheological behaviors (such as shear thickening \cite{FLBBO10,HTG17,GGGChG24} or the appearance of normal stress differences \cite{GMD16}) that do not occur in ordinary fluids. Given that suspensions are ubiquitous in nature and industry, from landslides and sediment transport \cite{MN06,PPS18,RMHBSG21} to food processing and civil engineering \cite{DM14,MT14,M20}, understanding their behavior is of great importance.

Among the different types of gas-solid flows, an interesting case is that of so-called particle-laden suspensions \cite{G94,J00,KH01,TS14,FH17,W20,LGW20}. In this type of suspension, diluted and immiscible particles are dispersed in a denser fluid, so hydrodynamic interactions lose relevance \cite{B72}. Instead, the dynamics of solid particles are mainly governed by thermal fluctuations in the fluid, in which interparticle, external, and Brownian forces predominate \cite{BB88}. Furthermore, it can often be assumed that the number density of the solid phase is much lower than that of the surrounding fluid (or solvent), allowing the latter to be treated as a thermostat at a fixed temperature. 

Under rapid-flow conditions, where particles interact primarily through binary, inelastic collisions and remain in contact only for short times, the system behaves as a granular gas. In this dilute regime, where interparticle collisions cannot be neglected \cite{S20}, the classical kinetic theory of gases \cite{CC70,FK72,RL77}, suitably modified to account for inelasticity and gas–solid coupling, provides an appropriate framework for modeling granular suspensions.

Despite their widespread relevance, the theoretical description of suspensions remains a major challenge. The coexistence of dissipative collisions between grains and the effects produced by the solvent, acting on different spatial and temporal scales, prevents the formulation of an exact description. In this context, the kinetic theory of gases, and in particular the Boltzmann and Enskog equations, continue to provide the most general theoretical framework for describing dilute and moderately dense granular suspensions, respectively, as they incorporate the essential ingredients associated with the characterization of granular gases, such as excluded volume and dissipation, without assuming equilibrium \cite{D00,NE01,L05a}. Due to the complexity of multiphase systems, a coarse-grained approach is often adopted where the effect of the solvent on grains is included through an effective force in the corresponding kinetic equation \cite{K90,G94,J00,K01}. Some models of solvent-particle flows describe this interaction by adding a viscous drag force proportional to the instantaneous particle velocity \cite{WGZS14,ASG19,SA20}. A more sophisticated model \cite{GTSH12} accounts for the thermal fluctuations of the solvent by adding a stochastic Langevin-like term, expressed in terms of the background temperature. This stochastic term represents the mean-field effect of random collisions between the solvent molecules and the grains, and is incorporated (along with the drag force term) into the kinetic equation through an additional Fokker--Planck term.

As the models described above for multiphase systems involve multiple approximations, their predictions should be carefully compared with numerical simulations. 
Even more, they also allow the exploration of regimes that are inaccessible experimentally. Simulations thus provide the necessary bridge between the microscopic properties of particles and macroscopic observables, offering a controlled environment for investigating the interaction between collisions and the effect of solvent on grains in granular suspensions.  

In practice, molecular dynamics (MD) and direct simulation Monte Carlo (DSMC) are among the most widely employed techniques for testing and validating the Boltzmann and Enskog equations. MD simulations provide a fully deterministic description of particle trajectories by solving Newton's equations of motion while adjusting the forces acting on the particles. This method is particularly effective for testing dense systems, where correlations and finite-size effects become important, and has been widely employed to test central hypotheses of kinetic theory, such as the behavior of correlation functions, the validity of the molecular chaos assumption, and the characterization of the velocity distribution in the inelastic regime. The accuracy of the Enskog equation has been systematically confirmed through comparisons with MD simulations \cite{LBD02,DHGD02,LLC07,MDCPH11,CS13,MGH14}, as well as with experimental data \cite{YHCMW02,HYCMW04}. This agreement holds particularly well for moderately dense systems, that is, for volume fractions $\phi \lesssim 0.2$ (where $\phi$ denotes the fraction of the total volume occupied by particles) and restitution coefficients $\alpha \gtrsim 0.8$ (with $\alpha$ being the normal coefficient of restitution that quantifies the degree of inelasticity of collisions). These analyses also extend to both monocomponent and polydisperse granular gases, highlighting the broad applicability of the Enskog framework \cite{GD99a,GD99b, L05, GDH07,GHD07}.  

On the other hand, the DSMC method, originally developed by Bird \cite{B63}, was specifically designed to address rarefied gas flows that lie beyond the computational reach of MD simulations. Like the Boltzmann equation, DSMC is based on the assumption of molecular chaos, treating collisions as stochastic and memoryless events, and its results remain fully consistent with Boltzmann's description even in the continuous regime \cite{B70b,B94}. Since its creation, DSMC has become a fundamental technique for validating and applying the Boltzmann equation to molecular gases, as well as for exploring fundamental questions such as the stability of its solutions and the transition to turbulence \cite{SC92,SC93,RN93,AG97,SC98}. Extensions of the Bird's method that incorporate a density-dependent collision enhancement factor, $\chi$, have enabled accurate simulation of moderate to high density regimes and have reproduced Enskog's predictions for uniform shear flow, pressure tensor, energy balance, and shear viscosity in molecular (elastic) fluids \cite{MS96,MS97}. Furthermore, DSMC studies have revealed that, in the high-velocity region, the distribution functions of granular gases exhibit exponential rather than Gaussian tails \cite{EP97,NE98,MG02,EB02a,EB02b}, a feature confirmed by simulations \cite{BCR99,MG02} and microgravity experiments \cite{YSS20}. Finally, DSMC has also provided detailed insights into the breakdown of energy equipartition in granular mixtures across a wide range of densities, offering strong support for the validity of the Enskog equation under such conditions \cite{ChGG22}.

Recently, Ref.~\cite{GABG23} derived analytical expressions for the granular tracer diffusion coefficient in the presence of a solvent from two complementary theoretical perspectives: the Enskog kinetic theory and a random-walk approach. Those results were compared with DSMC simulations, showing very good overall agreement and providing indirect support for the theoretical framework. However, that previous study was primarily focused on the analytical development of the theory and its internal consistency.

Motivated by these findings, the present work goes a significant step further by performing a systematic numerical validation of the theory. In particular, we revisit the problem of tracer diffusion by incorporating, in addition to DSMC data, results from Brownian molecular dynamics (MD) simulations. This combined DSMC–MD strategy allows us to directly test the validity of the Enskog–Langevin kinetic theory beyond the level explored in Ref.~\cite{GABG23}.

Our goal is to determine whether, and to what extent, the robustness of the Enskog description previously demonstrated for the \emph{dry} case (i.e., in the absence of an external fluid) can be extended to systems with viscous drag and stochastic forcing. Specifically, we aim to identify the range of friction parameters for which the kinetic description remains accurate, to analyze how the solvent alters velocity correlations and the velocity distribution function, and to quantify the overall reliability of the theory under these more realistic conditions. To clarify the origin of possible discrepancies between theory and MD simulations, our analysis is complemented with DSMC simulations, which provide an intermediate level of description between the theoretical approximations and the MD dynamics.

In this sense, the present work should be regarded as part of a progressive development of the kinetic theory of granular suspensions. It builds upon earlier studies on diffusion in dry sheared granular mixtures \cite{G02}, its extension to unsheared granular suspensions subject to drag and stochastic forcing \cite{GABG23}, and more recent generalizations to inertial suspensions under shear, where anisotropic diffusion tensors are introduced within the Langevin--Enskog framework \cite{THG23}.

With this aim, the remainder of the paper is organized as follows. Section II presents the theoretical background, including the Enskog equation for monocomponent systems, the suspension model, and its extension to mixtures. We then introduce the homogeneous steady states, which serve as the reference state for the Chapman--Enskog expansion used to compute the tracer diffusion coefficient. An alternative evaluation of the diffusion coefficient, based on the velocity correlation function, is also presented. Section III describes the computational methods employed, namely molecular dynamics (MD) and direct simulation Monte Carlo (DSMC) method. Section IV reports the results, focusing on the influence of the restitution coefficient and the intruder's mass, while Section V summarizes the main conclusions.

\section{Theoretical background}
\subsection{Kinetic Theory Foundations for Granular Gases}

At a kinetic level, all the relevant information on the microscopic state of the system is given by the one particle velocity distribution function $f(\mathbf{r}, \mathbf{v};t)$. This distribution function is defined such that $f(\mathbf{r},\mathbf{v};t)\dd\mathbf{r}\dd\mathbf{v}$ gives the average number of particles that, at time $t$, are within a volume element $\dd\mathbf{r}$ centered at position $\mathbf{r}$, and have velocities within the range $\dd\mathbf{v}$ around $\mathbf{v}$. Knowing the distribution function $f$ implies being able to link the macroscopic observables to the underlying microscopic dynamics.

The kinetic theory for inelastic hard spheres can be developed following steps similar to the elastic case. A rigorous derivation of the Boltzmann equation comes from a pseudo-Liouville formulation of the $N$ particle distribution function in the elastic case, which leads to the Bogoliubov--Born--Green--Kirkwood--Yvon (BBGKY) hierarchy relating the $n$ body to the $(n+1)$-body distribution functions \cite{C88,CIP13}. Since this hierarchy does not constitute a closed set of equations, a closure assumption is made for the two-particle distribution function $f_2(\mathbf{v}_2, \mathbf{r}_2, \mathbf{v}_1, \mathbf{r}_1; t)$, which denotes the joint probability of finding a particle in $\mathbf{r}_1$ with velocity $\mathbf{v}_1$, and another in $\mathbf{r}_2$ with velocity $\mathbf{v}_2$, at time $t$.

In the Boltzmann--Grad limit, suitable for dilute systems with short--range interactions, the BBGKY hierarchy reduces to the Boltzmann hierarchy \cite{NEB98,GST13,PS17}. Under these conditions, the molecular chaos hypothesis (or \emph{Stosszahlansatz}) postulates that
\beq\label{2.1}
f_2(\mathbf{r}_1, \mathbf{v}_1, \mathbf{r}_2, \mathbf{v}_2; t) \approx f(\mathbf{r}_1, \mathbf{v}_1; t) f(\mathbf{r}_2, \mathbf{v}_2; t)
\eeq
for pre-collision velocities. This factorization is valid before collisions, where velocity correlations are negligible. After interactions, however, particles develop velocity correlations, which are nevertheless negligible in dilute gases, since the probability of recollisions is minimal at significant time intervals, making the molecular chaos hypothesis a valid approximation.

While the Boltzmann equation applies to dilute gases, corrections are required at higher densities due to increased spatial correlations and excluded volume effects. In the regime of moderate densities, the semi-phenomenological revised Enskog kinetic theory (RET) provides a more accurate framework for describing dense elastic gases \cite{BE73a,BE73b,BE73c,BE79}. Enskog's approach takes into account the finite particle size by recognizing that the separation between the centers of colliding spheres cannot be neglected when the mean free path $\ell$ is comparable to the particle diameter $\sigma$. As a result,  the distribution function experiences spatial variation over distances $\sim\sigma$ during collisions. Although RET is still based on the assumption of molecular chaos, it incorporates spatial correlations through the pair correlation function at contact, $\chi[\mathbf{r}_1, \mathbf{r}_2 | n(\mathbf{r};t)]$. Here, $n(\mathbf{r};t)$ denotes the local number density, defined as
\beq\label{2.5}
n(\mathbf{r}; t) = \int \dd \mathbf{v} \, f(\mathbf{r}, \mathbf{v}; t).
\eeq
Hence, $\chi$ accounts for the increased probability of finding a pair of particles at contact due to finite-density effects and is the same functional of density as in a fluid in non-uniform equilibrium. Specifically, for particles that are about to collide, the two-body distribution function is approximated by
\beq\label{2.3}
f_2(\mathbf{r}_1, \mathbf{v}_1, \mathbf{r}_2, \mathbf{v}_2; t) \approx 
\chi(\mathbf{r}_1, \mathbf{r}_2 | n(\mathbf{r};t)) 
f(\mathbf{r}_1, \mathbf{v}_1; t)\, f(\mathbf{r}_2, \mathbf{v}_2; t).
\eeq
In the case of hard spheres undergoing instantaneous binary collisions, evaluated at contact (i.e., $\mathbf{r}_2 = \mathbf{r}_1 + \boldsymbol{\sigma}$, where $\widehat{\boldsymbol{\sigma}}$ is a unit vector along the line of centers of the two spheres at contact), the approximation \eqref{2.3} becomes
\beqa\label{2.4}
f_2(\mathbf{r}_1, \mathbf{v}_1, \mathbf{r}_1 + \boldsymbol{\sigma}, \mathbf{v}_2; t)
&\approx& 
\chi(\mathbf{r}_1, \mathbf{r}_1 + \boldsymbol{\sigma} | n(\mathbf{r};t)) \nonumber\\
&\times&
f(\mathbf{r}_1, \mathbf{v}_1; t)\,
f(\mathbf{r}_1 + \boldsymbol{\sigma}, \mathbf{v}_2; t).
\eeqa

\subsection{Enskog equation for a monocomponent granular gas}

Building on the kinetic description developed for a molecular gas, we now extend the analysis to a granular system where energy dissipation due to inelastic collisions plays a central role. Let us first consider a system consisting of a set of granular particles (grains) modeled as smooth inelastic hard particles, either disks in two dimensions ($d=2$) or spheres in three dimensions ($d=3$). Each particle has a mass $m$ and a diameter $\sigma$. The inelastic nature of the collisions is described only by a constant and positive coefficient of restitution $\alpha$, with $\alpha \le 1$. The assumption of perfectly smooth particles ensures that rotational degrees of freedom play no role, and energy dissipation occurs only through normal restitution. Under these conditions, the Enskog equation has the form

\beq
\label{3.1}
\frac{\partial f}{\partial t}+\mathbf{v}\cdot \nabla f+\frac{1}{m}\frac{\partial}{\partial \mathbf{v}}\cdot\left(\mathbf{F}f\right)=J_\text{E}[\mathbf{r},\mathbf{v}|f,f],
\eeq
where $\mathbf{F}(\mathbf{r},\mathbf{v};t)$ represents the action of an \textit{external} force (for instance, the gravity), and the Enskog collision operator is \cite[]{G19}
\begin{widetext}
\beqa
\label{3.2}
J_\text{E}[\mathbf{r},\mathbf{v}_1|f,f]&=&\sigma^{d-1}\int \dd\mathbf{v}_{2}\int \dd\widehat{\boldsymbol {\sigma
}}\ \Theta (\widehat{{\boldsymbol {\sigma }}} \cdot {\mathbf g}_{12}) (\widehat{\boldsymbol {\sigma }}\cdot {\mathbf g}_{12})
\left[\al^{-2}\chi(\mathbf{r},\mathbf{r}-\boldsymbol{\sigma})f(\mathbf{r},{\mathbf v}_{1}'';t)\right.\nonumber \\
& & \left. \times  f(\mathbf{r}-\boldsymbol{\sigma},{\mathbf v}_{2}'';t)-\chi(\mathbf{r},\mathbf{r}+\boldsymbol{\sigma})f(\mathbf{r},{\mathbf v}_{1};t)f(\mathbf{r}+\boldsymbol{\sigma},{\mathbf v}_{2};t)\right].
\eeqa
\end{widetext}
Here, $\mathbf{g}_{12}=\mathbf{v}_1-\mathbf{v}_2$ is the relative velocity, and $\Theta$ is the Heaviside step function. The relationship between the pre-collisional velocities $(\mathbf{v}_1'',\mathbf{v}_2'')$ and the post-collisional velocities $(\mathbf{v}_1,\mathbf{v}_2)$ is
\beqa
\label{3.3}
\mathbf{v}_1''&=&\mathbf{v}_1-\frac{1+\al}{2\al}(\widehat{{\boldsymbol {\sigma }}} \cdot {\mathbf g}_{12})\widehat{\boldsymbol {\sigma }}, \nonumber \\
\mathbf{v}_2''&=&\mathbf{v}_2+\frac{1+\al}{2\al}(\widehat{{\boldsymbol {\sigma }}} \cdot {\mathbf g}_{12})\widehat{\boldsymbol {\sigma }}.
\eeqa
Equations \eqref{3.3} characterize the \emph{restituting} collisions in which particles exit with certain final velocities. Reversing this process yields the standard formulation for direct collisions, where initial velocities $(\mathbf{v}_1, \mathbf{v}_2)$ evolve to final states $(\mathbf{v}_1', \mathbf{v}_2')$ as a result of collision dynamics \cite{BP04}:
\beqa
\label{3.4}
\mathbf{v}_1'&=&\mathbf{v}_1-\frac{1}{2}\left(1+\alpha\right)\left(\boldsymbol{\widehat{\sigma}}
\cdot\mathbf{g}_{12}\right)\boldsymbol{\widehat{\sigma}}, \nonumber \\ \mathbf{v}_2'&=&\mathbf{v}_2+\frac{1}{2}\left(1+\alpha\right)\left(\boldsymbol{\widehat{\sigma}}
\cdot\mathbf{g}_{12}\right)\boldsymbol{\widehat{\sigma}}.
\eeqa

Although the generalization of the Enskog equation to inelastic collisions constitutes a unique basis for the description of
granular gases at moderate densities, inelasticity in collisions affects both the velocity distribution and spatial structure of the system, leading to phenomena such as clustering, non-Gaussian velocity distributions, and breakdown of energy equipartition \cite{BP04,D07,DPS16,G19}. These effects become increasingly significant as density grows, potentially decreasing the regime where inelastic version of the RET remains accurate. In particular, the theory may fail to capture multi-particle correlations, long-range velocity correlations, or enduring contacts in very dense granular flows.

\subsection{Suspension model}
Let us now consider a system in which the granular particles are immersed in a carrier fluid (or solvent) that acts as a bath with viscosity $\eta_\text{b}$.  The influence of the solvent on the grain dynamics is modeled through an \emph{effective} external force acting on each grain \cite{KH01,GTSH12,GFHY16,GGG19a}. Consequently, the equation of motion for a particle with velocity $\mathbf{v}$ reads \cite{PLMPV98,NETP99,KSZ10,PGGSV12,KIB14}
\beq
\label{4.1}
m\dot{\mathbf{v}} = \mathbf{F}^\text{b} + \mathbf{F}^\text{coll},
\eeq
where $\mathbf{F}^\text{coll}$ accounts for interparticle collisions, while $\mathbf{F}^\text{b}$ represents the mean effect of the bath on the particle.

Equation\ \eqref{4.1} assumes that the effect of the solvent is \emph{decoupled} from that of collisions. This decoupling holds when the mean free time between collisions is 
much smaller than
the characteristic time over which the fluid significantly alters the grains' motion \cite{K90,TK95,SMTK96,KH01}. Under these conditions, the fluid can be considered sufficiently dilute, so that the Enskog collision operator remains identical to that of a dry (i.e., without external phase) granular system, and binary collisions mimic the mechanisms as if the solvent was absent. Moreover, we assume that the granular gas is dilute compared to the solvent, so that the latter remains essentially unaffected by the particles and acts as a thermostat at a given temperature $T_\text{b}$.

At low Reynolds numbers, the solvent-particle interaction $\mathbf{F}^\text{b}$ can be modeled by two independent contributions: (i) a viscous drag force and (ii) a stochastic Langevin-like force. The drag force accounts for friction with the fluid and is given by
\beq\label{4.2}
\mathbf{F}^\text{drag} = -m\gamma(\mathbf{v} - \mathbf{U}_\text{b}),
\eeq
where $\gamma$ is the drag coefficient and $\mathbf{U}_\text{b}$ is the mean velocity of the solvent. In the Enskog equation, this force is represented by the operator
\beq\label{4.3}
\mathbf{F}^\text{drag} f = -\gamma \frac{\partial}{\partial \mathbf{v}} \cdot (\mathbf{v} - \mathbf{U}_\text{b}) f.
\eeq
The stochastic force $\mathbf{F}^\text{st}$ models random collisions with fluid molecules and is represented as Gaussian white noise \cite{K81,THSG20}:
\beq\label{4.4}
\langle \mathbf{F}^\text{st}_i(t) \rangle = 0, \quad \langle \mathbf{F}^\text{st}_i(t) \mathbf{F}^\text{st}_j(t') \rangle = 2m^2\gamma T_\text{b} \mathrm{I} \delta_{ij} \delta(t - t'),
\eeq
where $T_\text{b}$ is the (constant) temperature of the background fluid and $\mathrm{I}$ is the identity tensor. The magnitude of the correlation is consistent with the fluctuation-dissipation theorem for elastic collisions. Following previous works \cite{GGG19a,GABG23}, the drag coefficient $\gamma$ is assumed to be a constant scalar quantity proportional to the solvent viscosity $\eta_\text{b}$ \cite{KH01}. 

In the kinetic equation, the stochastic force is represented by the Fokker–Planck operator \cite{NE98}\footnote{Throughout this work, temperature is expressed in energy units, i.e., $k_\text{B} = 1$.}:
\beq\label{4.5.1}
\mathbf{F}^\text{st} f = -\frac{\gamma T_\text{b}}{m} \frac{\partial^2}{\partial v^2} f.
\eeq

Taking into account all of these considerations, the Enskog equation \eqref{2.1} reads
\beq
\label{4.6}
\frac{\partial f}{\partial t}+\mathbf{v}\cdot \nabla f-\gamma\Delta\mathbf{U}\cdot\frac{\partial f}{\partial\mathbf{v}}-\gamma\frac{\partial }{\partial\mathbf{v}}\cdot(\mathbf{V} f)-\frac{\gamma T_\text{b}}{m}\frac{\partial^2 f}{\partial v^2}=J_\text{E}[\mathbf{r},\mathbf{v}|f,f].
\eeq
Here, $\Delta\mathbf{U}=\mathbf{U}-\mathbf{U}_\text{b}$, $\mathbf{V}=\mathbf{v}-\mathbf{U}$ is the peculiar velocity, and
\beq\label{4.7}
\mathbf{U}(\mathbf{r};t)=\frac{1}{n(\mathbf{r};t)}\int \dd\mathbf{v}\ \mathbf{v} f(\mathbf{r},\mathbf{v};t)
\eeq
is the mean flow velocity of the solid particles. 

The last relevant hydrodynamic quantity is the \emph{granular} temperature. Despite the energy loss due to inelastic collisions, numerous studies have shown that the granular temperature evolves on a slower timescale compared to other microscopic or kinetic quantities, allowing it to be treated as a \emph{slow} variable \cite{DHGD02}. It is defined as
\beq\label{4.8}
T(\mathbf{r};t) = \frac{m}{d n(\mathbf{r};t)} \int \dd\mathbf{v}\, \mathbf{V}^2(\mathbf{r};t) f(\mathbf{r}, \mathbf{v}; t),
\eeq

\noindent where $d$ is the dimensionality of the system. 
Since granular gases are inherently athermal systems, the granular temperature represents the variance of particle velocities relative to the local mean flow, and lacks a direct thermodynamic interpretation. However, when modeling particle-laden suspensions, the solvent %is typically 
can be characterized by a well-defined thermodynamic temperature $T_\text{b}$, which thermalizes the system. In such contexts, $T$ and $T_\text{b}$ can be formally treated on the same footing, as both quantify the intensity of velocity fluctuations \cite[]{G03}. This analogy justifies the use of Langevin-type models where fluctuation-dissipation relations are extended to granular flows.

In addition to revisiting the assumptions already tested in the dry case, the presence of a surrounding gas allows us to explore new aspects of the model using MD and DSMC simulations. These include, for instance, the decoupling between fluid forces and collisions, the validity of fluctuation-dissipation relations linking drag and noise, and the equality between the granular temperature $T$ and the bath temperature $T_\text{b}$ in the elastic limit. Among the different transport properties, the diffusion coefficient is particularly useful, as it captures the combined influence of collisions, drag, and stochastic forces. Comparing its theoretical prediction with simulation results provides a sensitive test of the kinetic theory and its reliability to model fluid-solid interactions. With this goal in mind, we now introduce a low number of impurities into the system.

\subsection{Mixture of intruders and grains}

In realistic situations, granular suspensions often exhibit polydispersity in the mechanical properties of its particles. To simplify the analysis and reduce the number of transport coefficients and parameters, we first consider a simplified system: a binary mixture of smooth inelastic hard disks or spheres, where one species is present in tracer concentration. The particles have masses $m$ and $m_0$, and diameters $\sigma$ and $\sigma_0$, with species $0$ representing a small number of impurities (or tracer particles).  In the tracer limit, collisions among impurities are negligible, so only impurity–grain interactions are considered. As before, inelastic collisions are characterized by a constant coefficient of normal restitution $\alpha_0 \leq 1$ for impurity–grain collisions.

Under these conditions, the Enskog kinetic equation for the granular gas distribution $f(\mathbf{r},\mathbf{v};t)$ decouples from the intruder distribution $f_0(\mathbf{r},\mathbf{v};t)$. Thus, $f$ obeys the closed Enskog equation \eqref{2.1} while the velocity distribution function $f_0$ for the intruder particles evolves under the influence of the granular background and the solvent. It satisfies the Enskog kinetic equation
\beq
\label{5.1}
\frac{\partial f_0}{\partial t}+\mathbf{v}\cdot \nabla f_0
- \gamma_0\frac{\partial}{\partial\mathbf{v}}\cdot\mathbf{v}f_0
- \frac{\gamma_0 T_{\text{b}}}{m_0}\frac{\partial^2 f_0}{\partial v^2}
= J_{0,\text{E}}[f_0,f],
\eeq
where the parameter $\gamma_0$ represents the drag coefficient for the intruder-solvent interaction which, like the drag coefficient $\gamma$, is assumed to be constant and treated as an input parameter. The collision term $J_{0,\text{E}}[f_0,f]$ describes intruder–grain interactions and is given by the Enskog–Lorentz operator \cite{G19}:
\beqa
\label{5.1.3}
J_{0,\text{E}}[\mathbf{r}_1,&  \mathbf{v}_1&|f_0,f] = \overline{\sigma}^{d-1} \int \dd\mathbf{v}_2 \int \dd\widehat{\boldsymbol{\sigma}}\, 
\Theta(\widehat{\boldsymbol{\sigma}} \cdot \mathbf{g}_{12}) 
(\widehat{\boldsymbol{\sigma}} \cdot \mathbf{g}_{12}) \nonumber\\
&\times& \Big[ \alpha_0^{-2} \chi_0(\mathbf{r}_1,\mathbf{r}_1 - \boldsymbol{\overline{\sigma}}) 
f_0(\mathbf{r}_1, \mathbf{v}_1'', t)\,
f(\mathbf{r}_1 - \boldsymbol{\overline{\sigma}}, \mathbf{v}_2'', t) \nonumber\\
&& - \chi_0(\mathbf{r}_1,\mathbf{r}_1 + \boldsymbol{\overline{\sigma}}) 
f_0(\mathbf{r}_1, \mathbf{v}_1, t)\,
f(\mathbf{r}_1 + \boldsymbol{\overline{\sigma}}, \mathbf{v}_2, t) \Big],
\eeqa
where $\chi_0$ is the intruder-granular particle pair correlation function at contact, $\overline{\sigma} = (\sigma + \sigma_0)/2$ is the average diameter, and $\boldsymbol{\overline{\sigma}} = \overline{\sigma} \widehat{\boldsymbol{\sigma}}$.

The pre-collisional (restituting) velocities $(\mathbf{v}_1'', \mathbf{v}_2'')$ are related to $(\mathbf{v}_1, \mathbf{v}_2)$ by
\beqa
\label{5.2}
\mathbf{v}_1'' &=& \mathbf{v}_1 - \mu (1 + \alpha_0^{-1}) (\widehat{\boldsymbol{\sigma}} \cdot \mathbf{g}_{12}) \widehat{\boldsymbol{\sigma}},\nonumber
\\
\mathbf{v}_2'' &=& \mathbf{v}_2 + \mu_0 (1 + \alpha_0^{-1}) (\widehat{\boldsymbol{\sigma}} \cdot \mathbf{g}_{12}) \widehat{\boldsymbol{\sigma}},
\eeqa
while the post-collisional velocities from direct collisions are given by
\beqa
\label{5.5}
\mathbf{v}_1'& = &\mathbf{v}_1 - \mu (1 + \alpha_0) (\widehat{\boldsymbol{\sigma}} \cdot \mathbf{g}_{12}) \widehat{\boldsymbol{\sigma}},\nonumber
\\
\mathbf{v}_2'& =& \mathbf{v}_2 + \mu_0 (1 + \alpha_0) (\widehat{\boldsymbol{\sigma}} \cdot \mathbf{g}_{12}) \widehat{\boldsymbol{\sigma}},
\eeqa
where the mass ratios are
\beq
\label{5.4}
\mu = \frac{m}{m + m_0}, \quad \mu_0 = \frac{m_0}{m + m_0}.
\eeq

\subsection{Homogeneous steady states}
\label{sec6}
In the long-time limit, the suspension tends toward a \textit{homogeneous} and \textit{stationary} state \cite{GGG21}, characterized by uniform  and time-independent hydrodynamic fields. In this regime, all spatial gradients vanish and the system reaches a steady balance between energy input from the bath and collisional energy dissipation.

The Enskog kinetic equations for the granular gas \eqref{3.1} and the intruder species \eqref{5.1} reduce to stationary and spatially homogeneous forms. Multiplying those equations by the kinetic energy per particle ($mv^2$ or $m_0 v^2$) and integrating over velocity leads to the steady-state energy balance equations. These take the form:

\beq
\label{6.1}
2\gamma \left(T_\text{b}-T \right)=T \zeta, \quad
2\gamma_0 \left(T_\text{b}-T_0 \right)=T_0 \zeta_0,
\eeq
where
\beq
T_0=\frac{1}{n_0 d} \int d\mathbf{v}\; m_0 v^2 f_0(\mathbf{v})
\eeq
is the intruder's temperature. Equation \eqref{6.1} expresses the energy balance in each subsystem: the stochastic bath injects energy via the noise term (proportional to $\gamma T_\text{b}$ or $\gamma_0 T_\text{b}$), while energy is dissipated through both the viscous drag term (proportional to $\gamma T$ or $\gamma_0 T_0$) and inelastic collisions, through the cooling rates $\zeta$ and $\zeta_0$, which represent the rate of kinetic energy loss due to grain-grain and intruder–grain collisions, respectively. They are defined as
\beqa
\label{6.3}
\zeta &=&-\frac{1}{n d T}\int d\mathbf{v}\; m v^2\, J_\text{E}[f,f], \nonumber \\
\zeta_0 &=&-\frac{1}{n_0 d T_0}\int d\mathbf{v}\; m_0 v^2\, J_{0,\text{E}}[f_0,f].
\eeqa

In the elastic limit ($\al = \al_0 = 1$), the cooling terms vanish ($\zeta = \zeta_0 = 0$), and the temperatures of both components equal the bath temperature: $T = T_0 = T_\text{b}$. In consequence, the corresponding velocity distributions are Maxwellians:
\beqa
\label{6.4}
f_\text{M}(\mathbf{v}) &=& n \left(\frac{m}{2\pi T_\text{b}}\right)^{d/2} \exp\left(-\frac{m v^2}{2T_\text{b}}\right), \nonumber\\
f_{0,\text{M}}(\mathbf{v}) &=& n_0 \left(\frac{m_0}{2\pi T_\text{b}}\right)^{d/2} \exp\left(-\frac{m_0 v^2}{2T_\text{b}}\right).
\eeqa

In the inelastic case, the velocity distributions $f$ and $f_0$ are not exactly known. Nonetheless, previous results \cite{GGG19a, GGG21} show that replacing them by Maxwellians at the steady temperatures $T$ and $T_0$ yields accurate estimates for the cooling rates. Under this approximation, the cooling rate for grains becomes
\beq
\label{6.6}
\zeta = \frac{1-\al^2}{d}\, \nu,
\eeq
where the effective collision frequency $\nu$ is
\beq
\label{6.7}
\nu = \frac{2^{d-1/2} d}{\sqrt{\pi}}\, \frac{\phi}{\sigma} \, \chi\, v_\text{th}, \quad 
v_\text{th} = \sqrt{\frac{2T}{m}},
\eeq
with
\beq
\label{6.7.2}
\phi=\frac{\pi^{d/2}}{2^{d-1}d\Gamma \left(\frac{d}{2}\right)}n\sigma^d
\eeq
the solid volume fraction.

For the intruders, the (reduced) partial cooling rate $\zeta_0^* = \zeta_0/\nu$ reads \cite{G19}
\begin{multline}
\label{6.8}
\zeta_0^* = \frac{2\sqrt{2}}{d}\, \mu \frac{\chi_0}{\chi} \left(\frac{\overline{\sigma}}{\sigma}\right)^{d-1} \left(\frac{1+\beta}{\beta}\right)^{1/2}(1+\al_0) \\
\times \left[1 - \frac{1}{2}\mu (1+\beta)(1+\al_0)\right],
\end{multline}
where 
\beq
\label{6.9}
\beta = \frac{m_0 T}{m T_0}
\eeq
is the ratio of mean square velocities between intruders and gas particles.

Equations \eqref{6.1}, \eqref{6.6}, and \eqref{6.8} form a closed set that determines the steady temperatures $T$ and $T_0$. Despite its simplicity, the Maxwellian approximation shows excellent agreement with simulation data over a wide parameter range \cite{GABG23}, which justifies its use in theoretical predictions. In particular, the theoretical predictions for the steady temperatures $T$ and $T_0$ have been shown to agree remarkably well with DSMC simulations \cite{GABG23}. Moreover, previous work \cite{GGG21} has analyzed the deviations of the distribution functions from their Maxwellian forms by computing the fourth-order cumulants of the velocity distribution functions. It was observed that, in the presence of an external fluid, these cumulants are significantly smaller than in the dry granular case, indicating that the distributions are closer to Maxwellians. Physically, this stems from the fact that the external driving (white noise) tends to thermalize the system and suppress high-velocity tails, in contrast to the freely cooling dry case where such tails are enhanced by inelasticity.

In the present work, the theoretical predictions based on the Maxwellian approximation will be further tested against MD simulations over a broad range of values for the drag coefficients $\gamma$ and $\gamma_0$. This allows us to assess the robustness of the approximation not only with respect to inelasticity, but also with respect to the strength of the solvent-particle force.

\subsection{Tracer diffusion coefficient}

In this section, we summarize the calculations carried out in Ref.~\cite{GABG23} for the evaluation of the tracer diffusion coefficient $D$ in a granular suspension. It characterizes the mean squared displacement (MSD) of an intruder immersed in a granular gas interacting with an solvent, in the limit of weak concentration gradients. Since intruders may freely exchange momentum and energy with the grains, only their number density $n_0$ is conserved. In this regime, and under the assumption of a hydrodynamic description, the intruder number density $n_0$ satisfies a diffusion equation:
\beq
\label{7.2}
\frac{\partial n_0}{\partial t} = D \nabla^2 n_0,
\eeq
with the corresponding Einstein relation for the MSD:
\beq
\label{7.3}
\langle |\Delta \mathbf{r}(t)|^2 \rangle = 2 d D t.
\eeq

To compute $D$, the Chapman--Enskog (CE) \cite{CC70} expansion is applied to the Enskog--Lorentz kinetic equation for the intruder, assuming a normal solution and expanding the distribution function $f_0$ in powers of the density gradient $\nabla n_0$. The assumption of a normal solution implies that all space and time dependence of $f_0$ occurs only through the hydrodynamic fields, which evolve on slow (macroscopic) time scales. This allows a systematic perturbative treatment where transport coefficients, such as the tracer diffusion coefficient, appear as leading-order contributions in the gradient expansion. The resulting integral equation for the first-order correction $f_0^{(1)}$ is approximately solved by considering the two first Sonine polynomials corrections:
\beq
\label{7.4}
\boldsymbol{\mathcal A}(\mathbf{v}) \simeq -f_{0,\text{M}}(\mathbf{v})\Big[a_1 \mathbf{v} + a_2 \mathbf{S}_0(\mathbf{v})\Big],
\eeq
where $f_{0,\text{M}}$ is a Maxwellian at the intruder temperature $T_0$ and  
\beq
\label{7.5}
\mathbf{S}_0(\mathbf{v})=\Big(\frac{1}{2}m_0 v^2-\frac{d+2}{2}T_0\Big)\mathbf{v}
\eeq
is the second Sonine polynomial. The first and second Sonine approximations correspond to retaining only $a_1$ or both $a_1$ and $a_2$, respectively.
Under these assumptions, the first Sonine approximation to the diffusion coefficient $D$ is given by \cite{GABG23}
\beq
\label{7.6}
D[1] = \frac{T_0/m_0}{\gamma_0 + \nu_a}.
\eeq
In addition, the second Sonine approximation improves the convergence of the expansion by including higher-order terms by means of a correction factor $\Lambda$, leading to the expression
\beq
\label{7.8}
D[2] = \Lambda D[1],
\eeq
where
\beq
\label{7.9}
\Lambda = \frac{(\nu_a + \gamma_0)(\nu_d + 3\gamma_0)}{(\nu_a + \gamma_0)(\nu_d + 3\gamma_0) - \nu_b \left[ \nu_c + 2\gamma_0 \left(1 - \frac{T_\text{b}}{T_0} \right) \right]}.
\eeq
The collision frequencies $\nu_a$, $\nu_b$, $\nu_c$, and $\nu_d$ are given in Appendix \ref{appA}.

While this approach yields analytical expressions for $D$, the accuracy of the CE expansion (especially at moderate-to-strong dissipation) must be assessed. For this reason, the theoretical predictions are compared against DSMC and MD simulations throughout this work. These simulations provide independent estimates of $D$ from the long-time behavior of the intruder MSD via Eq.\ \eqref{7.3}, serving as a test of the evolution of $f_0$ toward a normal solution in which all time dependence occurs solely through the granular temperature $T$, as well as a validation of the Sonine approximations.

\subsection{Velocity correlation function}
The Green--Kubo formalism provides an exact relation between transport coefficients and time correlation functions of the corresponding fluxes in equilibrium or non-equilibrium steady states \cite{GN00,DBL02,BR04}. In this formalism, the tracer diffusion coefficient $D$ is given by
\beq \label{8.1}
D = \frac{1}{d} \int_0^\infty \dd t \, \langle \mathbf{v}(t) \cdot \mathbf{v}(0) \rangle \equiv \int_0^\infty \dd t \, C_{vv}(t),
\eeq
where
\beq
C_{vv}(t) = \frac{1}{d} \langle \mathbf{v}(t) \cdot \mathbf{v}(0) \rangle
\eeq
is the velocity autocorrelation function (VACF).

The intruder evolves following the Langevin equation \eqref{4.1}, with mass $m_0$ and friction coefficient $\gamma_0$. The collisional force $\mathbf{F}^\text{coll}(t)$ arises from inelastic collisions with grains and is responsible for additional dissipation and momentum transfer. If the response functions are expanded to
leading order in a cumulant expansion \cite{DG01}, this term can be approximated as an effective friction
\begin{equation}
\mathbf{F}^\text{coll}(t) \approx -\nu_a \mathbf{v}(t),
\end{equation}
where $\nu_a$ is an effective collision frequency given by
\beq
\nu_a = - \frac{m_0}{d n_0 T_0} \int \dd\mathbf{v} \, \mathbf{v} \cdot J_{0,\text{E}}[f_{0,\text{M}}\mathbf{v},f].
\eeq

In this approximation, the equation of motion becomes:
\begin{equation}
\frac{d\mathbf{v}}{dt} = - (\gamma_0 + \nu_a) \mathbf{v} + \frac{\mathbf{F}^\text{st}(t)}{m}.
\end{equation}

\noindent where $\mathbf{F}^{\text{st}}(t)$ is the Gaussian white noise, decorrelated in time and defined in Eq.~\eqref{4.4}, and $\gamma_0$ is the drag coefficient associated with the intruder--solvent interaction. This is a classical Ornstein--Uhlenbeck process with total damping $\gamma_\text{eff} = \gamma_0 + \nu_a$. The solution of the velocity under this process is
\begin{equation}
\mathbf{v}(t) = \mathbf{v}(0) e^{-\gamma_\text{eff} t} + \frac{1}{m}\int_0^t \dd s\, e^{-\gamma_\text{eff} (t - s)} \mathbf{F}^\text{st}(s) .
\end{equation}
Taking the average with $\mathbf{v}(0)$ and using the independence of noise and initial condition
\begin{equation}\label{ct}
 C_{vv}(t) = \frac{1}{d}\langle \mathbf{v}(0)^2 \rangle e^{-\gamma_\text{eff} t} = C_{vv}(0) e^{-\gamma_\text{eff} t},
\end{equation}
where
\begin{equation}
C_{vv}(0) = \frac{1}{d}\langle v^2 \rangle = \frac{ T_0}{m_0},
\end{equation}
and $T_0$ is the steady-state kinetic temperature of the intruder.
Introducing the dimensionless scaled time
\beq
\tau = \frac{\gamma_0}{m} t,
\eeq
the VACF becomes
\beq\label{8.2}
C_{vv}(\tau) = C_{vv}(0) \, \exp \left[ - \left( m + \frac{m \nu_a}{\gamma_0} \right) \tau \right].
\eeq
Finally, inserting Eq.\ \eqref{ct} into Eq.\ \eqref{8.1}, the diffusion coefficient becomes
\beq
D = \frac{T_0/m_0}{\gamma_0 + \nu_a},
\eeq
which coincides with the first Sonine approximation obtained by the Chapman--Enskog method.

\section{Computational methods for dilute molecular gases}

%\section{Simulation model}

Simulations have been performed to test the theoretical predictions in several cases: Molecular dynamics simulations in a system of spherical particles with a single intruder, or tracer, and Monte Carlo simulations to solve numerically the Enskog equation.

\subsection{Molecular dynamics}

Simulations are performed in a three dimensional system of $N=1000$ hard spherical particles, obeying the microscopic Eq.\ \eqref{4.1}, which we rewrite explicitly for particle $j$ as ($j=0$ denotes the intruder):

\begin{equation}
m_j \frac{d^2 \rb_j}{dt^2} = -m_j \gamma_j \left(\vb_j-\mathbf{U}_{\text{b}}\right) + \mathbf{F}^\text{st} + \mathbf{F}^\text{coll} \label{Langevin}
\end{equation}

All particles have the same diameter $\sigma$, whereas the mass of the intruder, $m_0$, has been varied, with the rest of particles having the same mass $m$. The friction coefficient with the fluid, $\gamma$, has been set a small value, as further discussed bellow, to accomplish for the condition imposed in the theory; furthermore, $\gamma_j=\gamma$ for $j\neq 0$. %This also applies to the stochastic for $\mathbf{F}^\text{st}$, which depends of the value of $\gamma_j$. 
Equation \eqref{Langevin} is integrated, without collision forces, using a Heun algorithm in small time steps, $\delta t$. After every integration step, overlaps between neighboring particles are checked, and collisions are treated using Eq.\ \eqref{5.5}. 

In the simulations, the diameter of the particles, the mass of the bath particles and the thermal energy have been set equal to $1$. The ratio of the intruder mass, $m_0/m$ has been varied between $10^{-2}$ and $10^2$. The particle density, reported as the volume fraction, has been kept low to comply with the conditions of the theory. Periodic boundary conditions have been  used in all three dimensions, the center of mass of the system was not fixed, and in all cases $U_\text{b}=0$. The time step for the integration of the equations of motion is $2.5\cdot 10^{-3} \sigma \sqrt{m/T_{\text{b}}}$ for $m_0>0.1m$ and $2.5\cdot 10^{-4} \sigma \sqrt{m/T_{\text{b}}}$ for $m_0\leq0.1m$ (recall that the Boltzmann constant has been set to $1$ and $T_{\text{b}}$ has units of energy). In selected cases, simulations with $N=8000$ particles have been run, as shown below, to test for finite size effects. All relevant parameters of the simulations are given in Table \ref{parameters}.

\begin{figure}
    \includegraphics[width=\columnwidth]{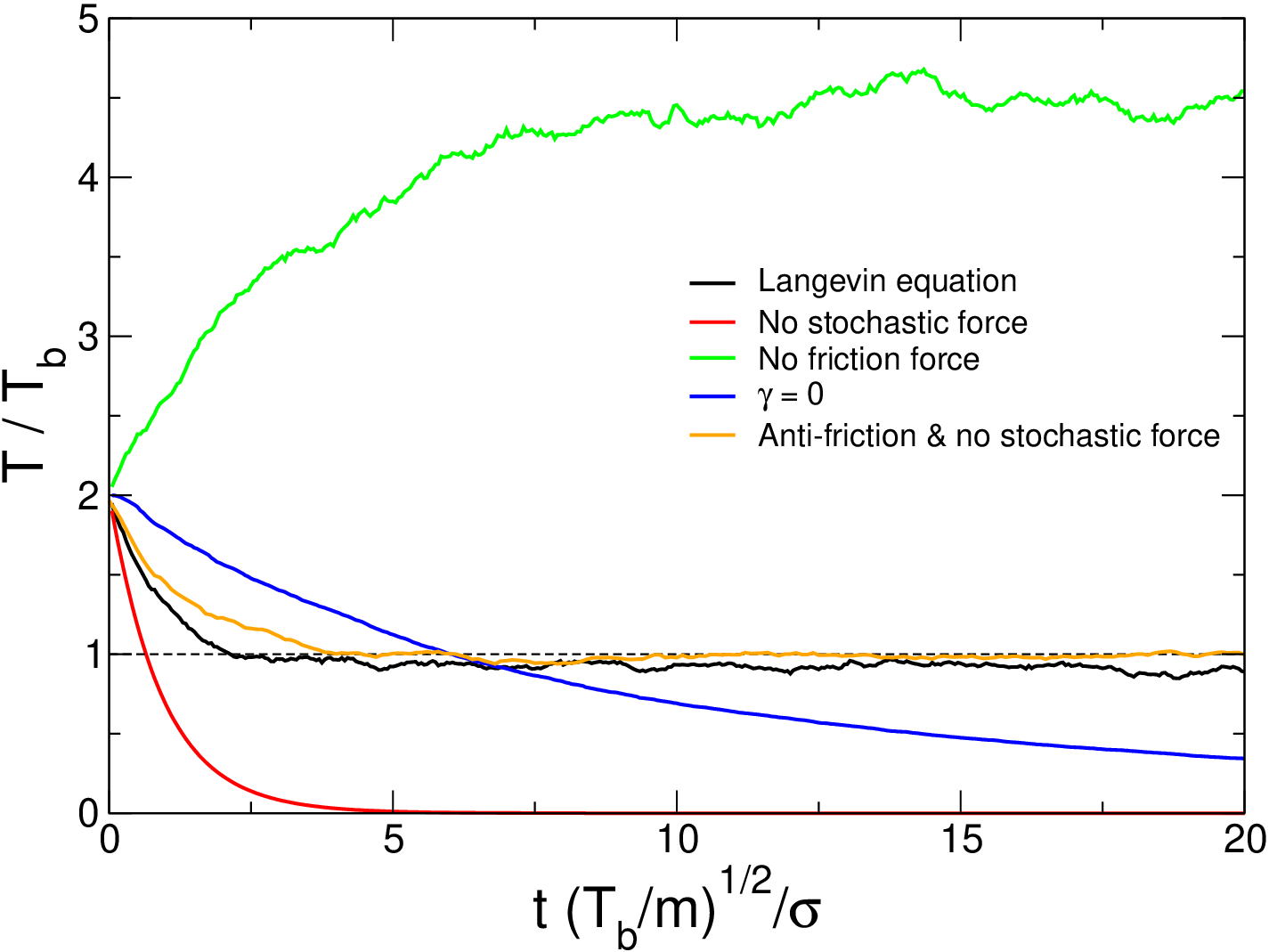}
    \caption{Evolution of the average kinetic energy, $K$, per particle following Eq.\ \eqref{Langevin} (black line) from MD simulations. The volume fraction is $\phi=0.10$ and  restitution coefficient $\alpha=0.8$. The intruder has the same properties as the bath particles.}
    \label{equilibration}
\end{figure}

The performance of the Langevin equation is presented in Fig. \ref{equilibration}, which shows the effect of the stochastic and friction forces in the thermalization of the system with $\alpha =0.8$. For this purpose, the evolution of the kinetic energy per particle in a single evolution is presented, starting from a configuration  above the equilibrium value (horizontal dashed line). The blue line corresponds to the absence of solvent, featuring the homogeneous cooling state, whereas the black line shows the evolution following the full Langevin equation. In the latter, the kinetic energy reaches a steady value slightly below $T_{\text{b}}$, due to the inelastic collisions. The red and green lines show the sole contributions of the friction and stochastic forces, respectively. While the drag force, first term in Eq.\ \eqref{Langevin}, describes the dissipation of the energy, and therefore the kinetic energy vanishes, the stochastic term injects energy well above the equilibrium value. Both contributions balance as stated by the fluctuation-dissipation theorem, yielding a steady state, matching the equilibrium value only for elastic collisions. Inelastic collisions imply an extra loss of energy which the stochastic force balances at a lower, but constant, value of the kinetic energy.

An alternative thermostat can be constructed with a negative friction force, without the stochastic contribution. This injects energy in the system to balance the dissipation in collisions, and the system reaches a steady state at long times. The (negative) value of $\gamma$ sets the kinetic energy in the steady state. In the figure, the orange line represents this case with $\gamma=-0.1 \sqrt{m T_{\text{b}}}/\sigma$, which reaches a stationary state comparable to the Langevin thermostat (note that $T_{\text{b}}$ does not appear in the equation of motion since there is no stochastic force, but it represents the thermal energy according to equipartition in the elastic case). Although this thermostat produces an evolution of the kinetic energy similar to the Langevin equation, the latter is the physical model for a suspension and is therefore preferable.

\subsection{Direct Simulation Monte Carlo (DSMC) Method}

To simulate the dynamics of the granular suspension, we employ the Direct Simulation Monte Carlo (DSMC) method. 
This is a stochastic algorithm that provides numerical solutions to the Boltzmann or Enskog kinetic equations under the molecular chaos assumption \cite{B70b,B94,MG02}. It is particularly well-suited for rarefied systems where particle interactions are predominantly binary and uncorrelated. By simulating the velocity evolution of a set of representative particles, DSMC approximates the velocity distribution function $f_i(\mathbf{v}, t)$ of each species $i$ through an ensemble of $N_i$ simulated particles.

Initially, particle velocities are sampled from a prescribed distribution at a temperature $T_{i}^{(0)}$ for each species. The discrete approximation to the distribution function is constructed as \cite{MG02}
\beq
f_i^{(N)}(\mathbf{v}, t) \to \frac{n_i}{N_i} \sum_{k=1}^{N_i} \delta(\mathbf{v} - \mathbf{v}_k(t)),
\eeq
where $n_i$ is the number density, and $\mathbf{v}_k$ are the instantaneous velocities of the simulated particles of species $i$.

Since we consider a spatially homogeneous system, the simulation proceeds by alternating two stages: (i) binary collisions and  (ii) the action of the solvent.

\subsubsection{Grain--Grain Collisions}

The number of candidate collision pairs between granular particles in a time interval $\delta t$ is given by
\beq
\label{DSMC1}
N^{\delta t} = \frac{2^{d-1} d\, \Gamma\left(\frac{d}{2}\right)}{\pi^{(d-2)/2}} \frac{N}{ \sigma^{d-2}} \, \phi \chi g^{\text{max}} \delta t,
\eeq
where $N$ is the number of simulated granular particles. The value $g^{\text{max}}$ is an upper bound for the relative velocity, estimated as $g^{\text{max}} = C v_{\text{th}}$, with $v_{\text{th}} = \sqrt{2 T / m}$ and $C \sim 5$ \cite{B94}. The factor $\chi$ is included to enhance the number of candidate pairs in order to solve the Enskog equation \cite{MS97}.

A colliding direction $\widehat{\boldsymbol{\sigma}}$ is randomly sampled on the unit sphere (or unit circle in $d=2$), and a candidate pair $(k,\ell)$ undergoes a collision if
\beq
\left| \widehat{\boldsymbol{\sigma}} \cdot \mathbf{g}_{k\ell} \right| > \text{R}(0,1) \, g^{\text{max}},
\eeq
where $\mathbf{g}_{k\ell} = \mathbf{v}_k - \mathbf{v}_\ell$ is the relative velocity, and $\text{R}(0,1)$ is a uniform random number in $[0,1]$. If accepted, the post-collisional velocities are updated according to the scattering rules \eqref{3.4}.

\subsubsection{Intruder--Grain Collisions}

Intruders interact only with gas particles. The number of candidate intruder–gas collision pairs in $\delta t$ is
\beq
N_{0}^{\delta t} = \frac{2^{d-1} d\, \Gamma\left(\frac{d}{2}\right)}{\pi^{(d-2)/2}} \frac{N_0  \, \overline{\sigma}^2}{ \sigma^d} \, \phi \chi_{0} g_{0}^{\text{max}} \delta t,
\eeq
where  $g_{0}^\text{max}=C\sqrt{2T / \overline{m}}$ is an upper bound for the intruder–gas relative velocity, with $\overline{m}=(m+m_0)/2$. The collision is accepted with the same criterion as in the grain–grain case, and the velocities are updated following the binary collision rules \eqref{5.5}.

\subsubsection{Effect of the Solvent}

To model the interaction with the solvent, a stochastic Langevin-type thermostat is applied at each time step $\delta t$. In the three-dimensional case ($d=3$), the velocity of each particle of species $i$ is updated according to
\beq
\mathbf{v}_k \to e^{-\gamma_i \delta t} \mathbf{v}_k + \left( \frac{6 \gamma_i T_\text{b} \delta t}{m_i} \right)^{1/2} \mathbf{R},
\eeq
where $\mathbf{R}$ is a random vector uniformly distributed in $[-1,1]^3$, $T_\text{b}$ is the bath (gas) temperature, and $\gamma_i$ is the drag coefficient of species $i$. This stochastic update reproduces the action of a Fokker--Planck operator in the limit where the time step $\delta t$ is much smaller than the mean time between collisions \cite{KG14}.

\subsection{Measured Quantities}

A central goal of this study is to evaluate the accuracy of the Enskog kinetic theory in predicting the transport properties of an intruder immersed in a gas of granular particles. To this end, we focus on two key observables: the intruder's kinetic temperature and its tracer diffusion coefficient. These quantities are directly accessible through MD and DSMC simulations and admit explicit theoretical expressions derived from the Enskog kinetic equation. Their comparison enables a quantitative assessment of the validity of the assumptions used in the kinetic theory framework (such as the leading Sonine polynomial approximations, the adoption of Maxwellian velocity distributions, and the application of the Chapman--Enskog expansion to first order in spatial gradients) across a broad range of physical parameters.

On the one hand, we measure the instantaneous kinetic temperature of intruders. It is defined as
\beq
T_0(t) = \frac{m_0}{d} \left\langle v_i^2(t) \right\rangle,
\eeq
where $\langle \cdot \rangle$ denotes an average over all intruders at time $t$. As the system evolves, $T_0(t)$ eventually reaches a steady-state value.

On the other hand, the tracer diffusion coefficient $D$ of the intruder is computed using the Einstein relation:
\beq
\label{5.1}
D = \frac{1}{2d \delta t} \left[ \langle R^2(t + \delta t) \rangle - \langle R^2(t) \rangle \right],
\eeq
where $\langle R^2(t) \rangle$ is the mean-square displacement of the intruder particles at time $t$, averaged over the ensemble of $N_0$ simulated intruders. All statistical averages are taken in the steady state and over multiple independent realizations.

For the results presented below, the MD simulations were carried out over 500 independent trajectories of the system with length $t \gamma / m = 500$, in adimensional time units. The VACF is calculated from the particle velocities, which are calculated alongside positions in the Langevin equation with the Heun algorithm.

In the DSMC simulations, a total of $10^5$ intruder particles were simulated. Their dynamics do not influence the state of the surrounding granular gas. At each DSMC time step, an average of 200 candidate collision pairs is proposed, a number calibrated to reproduce the expected Boltzmann collision rate. For each set of parameters, we record 2.5$\times 10^4$ values of relevant observables (namely, the intruder temperature and mean-square displacement) distributed across 5 independent simulation runs.

\begin{widetext}
\begin{table*}
\centering
\begin{tabular}{lcr} \hline
Parameter & Description & Value / range \\ \hline
$d$ & dimensionality of the system &  3 \\
$\phi$ & Volume fraction & $0.05-0.20$ \\
$T_\text{b}$ & Bath temperature & 1 \\
$\gamma$ & Friction coefficient of granular gas & $1 \sqrt{mT_\text{b}}/\sigma$ \\
$\gamma_0$ & Friction coefficient of tracer particle & $\gamma_0=\gamma$ \\
$m$ & Mass of granular particle & 1\\
$m_0$ & Mass of tracer particle & $0.01m, \dots, 100m$ \\
$\sigma$ & Diameter of granular particle & 1\\
$\sigma_0$ & Diameter of tracer particle & $\sigma_0=\sigma$\\
$\alpha=\alpha_0$ & Coefficients of restitution & $0.3,\dots, 1$ \\ \hline
\end{tabular}
\caption{Parameters of the simulated systems.}
\label{parameters}
\end{table*}
\end{widetext}

\section{Results}
One of the key issues when working with suspensions is deciding what value to assign to the drag parameter, or more generally, to the term that models the action of the solvent. In the model considered here, two main assumptions are made: (i) the force or influence of the solvent must be weak enough for the Enskog collision operator to remain the same as in a dry gas, and (ii) collisions must not be the only mechanism controlling the movement of the grains, completely neglecting the effect of the solvent, as this would correspond to a Knudsen gas regime.

In previous works on granular suspensions \cite{HTG17,GGG19b}, the effect of the solvent is incorporated into a dimensionless parameter 
$\gamma^*$, defined as  
\begin{equation}
\gamma^* = \frac{\gamma}{n\sigma^{d-1}\sqrt{2T/m}}
= \frac{\sqrt{2}\,\pi^{d/2}}{2^d d\,\Gamma\!\left(\tfrac{d}{2}\right)}\, 
\frac{1}{\phi\,T_\text{b}^*} 
\left(\frac{T_\text{b}}{T}\right)^{1/2},
\end{equation}
where $n\sigma^{d-1}\sqrt{2T/m}$ characterizes the contribution from 
particle--particle collisions, while $\gamma$ accounts for the influence of the 
solvent. The parameter $T_\text{b}^* = T_\text{b}/(m\sigma^2\gamma^2)$ 
is introduced for convenience. For moderate solid volume fractions, 
$\phi \lesssim 0.3$, $T_\text{b}^*$ is typically chosen to be of order unity, 
ensuring that the effects of collisions and the solvent are of comparable magnitude.

As discussed in Ref.\ \cite{GABG23}, $T_\text{b}^*$ can be brought close to unity either by controlling the particle diameter $\sigma$ or mass $m$ or by selecting a molecular gas with a suitable viscosity $\eta_\text{b}$. For example, for a system composed of hydrogen as the molecular gas and gold grains, the Reynolds numbers remain of order $10^{-4}$ or less, and the Stokes numbers, $T_\text{b}^*$, and $\gamma/\nu$ are all close to unity. Moreover, variations in the granular temperature within reasonable bounds do not significantly affect these parameters, keeping them consistent with the hypotheses of the model.

To be more precise, we will compare the theoretical predictions for the tracer diffusion coefficient $D$ across different values of the drag parameter, with the aim of identifying the regimes in which the theoretical assumptions fail when contrasted with MD or DSMC results. For simplicity, all simulations are carried out with $\gamma_0 = \gamma$, hard spheres ($d=3$), and $\alpha_0 = \alpha$.

\begin{figure}
    \includegraphics[width=\columnwidth]{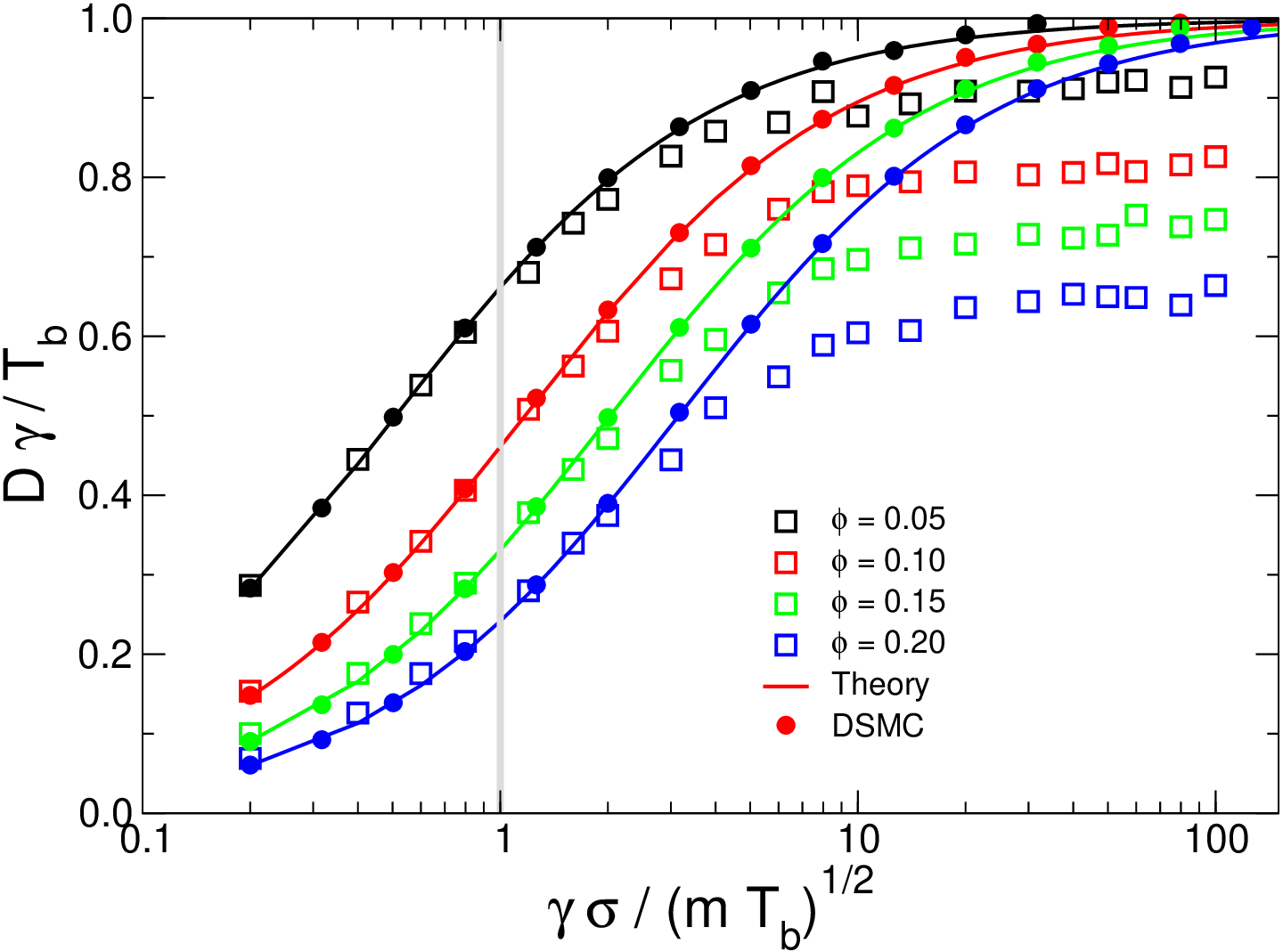}
    \caption{Tracer diffusion coefficient as a function of the friction coefficient $\gamma$ for different volume fractions, as labeled. Simulation data are shown as open circles, lines represent the theory and closed symbols indicate the DSMC calculations. Error bars are smaller than the symbol size.}
    \label{D-gamma0}
\end{figure}

Figure \ref{D-gamma0} shows the tracer diffusion coefficient as a function of the drag coefficient, $\gamma$, from simulations and theory, scaled with the bare (in absence of collisions) particle diffusion coefficient. Here, the intruder has the same mass as the bath particles and all collisions are elastic. For all volume fractions $D/(T_\text{b}\gamma)$ grows with $\gamma$ indicating the increasing relevance of Brownian forces in the diffusion of particles. Nevertheless, even in the high-$\gamma$ limit diffusion is affected by the collisions, which hinder it.

The theory (lines) and DSMC results (closed symbols) correctly captures the overall phenomenology, but fails in the high-$\gamma$ regime, as single particle diffusion is reached in this limit for all particle densities. This indicates that the effect of collisions on diffusion are neglected in the limit of large Brownian forces.  Notably, for moderate densities and low to intermediate $\gamma$, the agreement between theory and simulations is quantitative. This comparison also allows us to select a value of $\gamma$ where the theory matches the simulation values; in the following, we select $\gamma \sigma/\sqrt{m T_b}=1$. With this choice, $T_\text{b}^*=T_\text{b}/(m \sigma^2\gamma^2)=1$, indicating a balance between collisional effects and the influence of the solvent.

\subsection{Effect of the restitution coefficient}

\begin{figure}
    \includegraphics[width=\columnwidth]{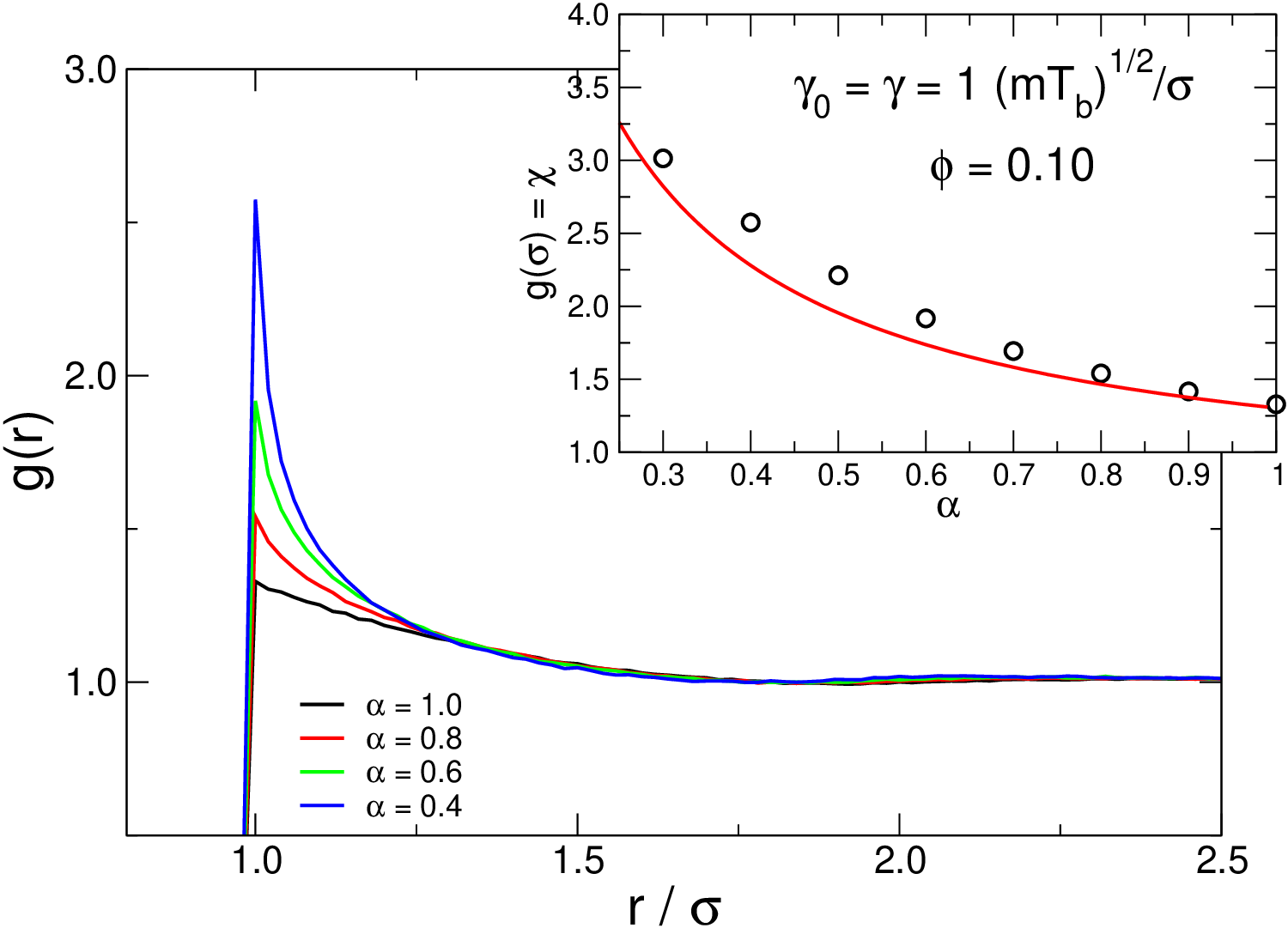}
    \caption{Pair distribution function of the system with volume fraction $\phi=0.10$ and friction coefficient $\gamma_0=\gamma=1 (m T_b)^{1/2}/\sigma$, from MD simulations for different restitution coefficients, as labeled. Inset: Contact value of the pair distribution function versus the restitution coefficient, $\alpha$, from theory (line) and MD simulations (open circles).}
    \label{gr}
\end{figure}

In this subsection, we study the effect of inelastic collisions, combined with the action of the surrounding gas, in the dynamics of the system. We analyze first the effect of $\alpha$ on the local structure of the system. In particular, the modification introduced in the Enskog equation to account for spatial correlations must be assessed. Figure \ref{gr} shows the radial distribution function $g(r)$ for several values of the restitution coefficient $\alpha$ at a volume fraction $\phi = 0.10$. The inset displays the contact value $\chi=g(\sigma)$ as a function of $\alpha$.
 As expected, inelastic collisions modify only the structure of the system at short distances, whereas for distances beyond $1.25\sigma$ the pair distribution function is unaffected. The value of the function at contact, $\chi$, which is incorporated into the theory and compared with the theoretical prediction obtained in \cite{L01} for the homogeneous cooling state (HCS), i.e., freely cooling granular gases:

\begin{equation}
\chi(\alpha) = \frac{1+\al}{2\al}\chi_\text{el}
\end{equation}

\noindent where $\chi_\text{el}$ is the contact value for elastic collisions, that can be taken from Carnahan--Starling \cite{CS69}:

\begin{equation}
\chi_\text{el} = \frac{1 - \tfrac{1}{2}\phi}{(1 - \phi)^3},
\end{equation}

The comparison shows excellent agreement for large values of $\alpha$, while noticeable deviations appear for $\alpha \lesssim 0.8$, consistent with the findings of Lutsko \cite{L01}. It should be noted, however, that Lutsko’s expression was derived for the homogeneous cooling state (HCS), whereas our results pertain to granular suspensions. Nevertheless, following previous works in the granular literature \cite{G19}, we retain the Carnahan--Starling expression as a reliable approximation, since Fig.\ \ref{gr} shows no significant deviations for moderate inelasticities. 

%while the intruder--gas pair correlation function is given by \cite{B70}
%\begin{equation}
%\chi_0 = \frac{1}{1-\phi}+3\frac{\omega}{1+\omega}\frac{\phi}{(1-\phi)^2}+2
%\frac{\omega^2}{(1+\omega)^2}\frac{\phi^2}{(1-\phi)^3},
%\end{equation}
%where $\omega = \sigma_0/\sigma$ denotes the diameter ratio between the intruder and grains.}

\begin{figure}
    \includegraphics[width=\columnwidth]{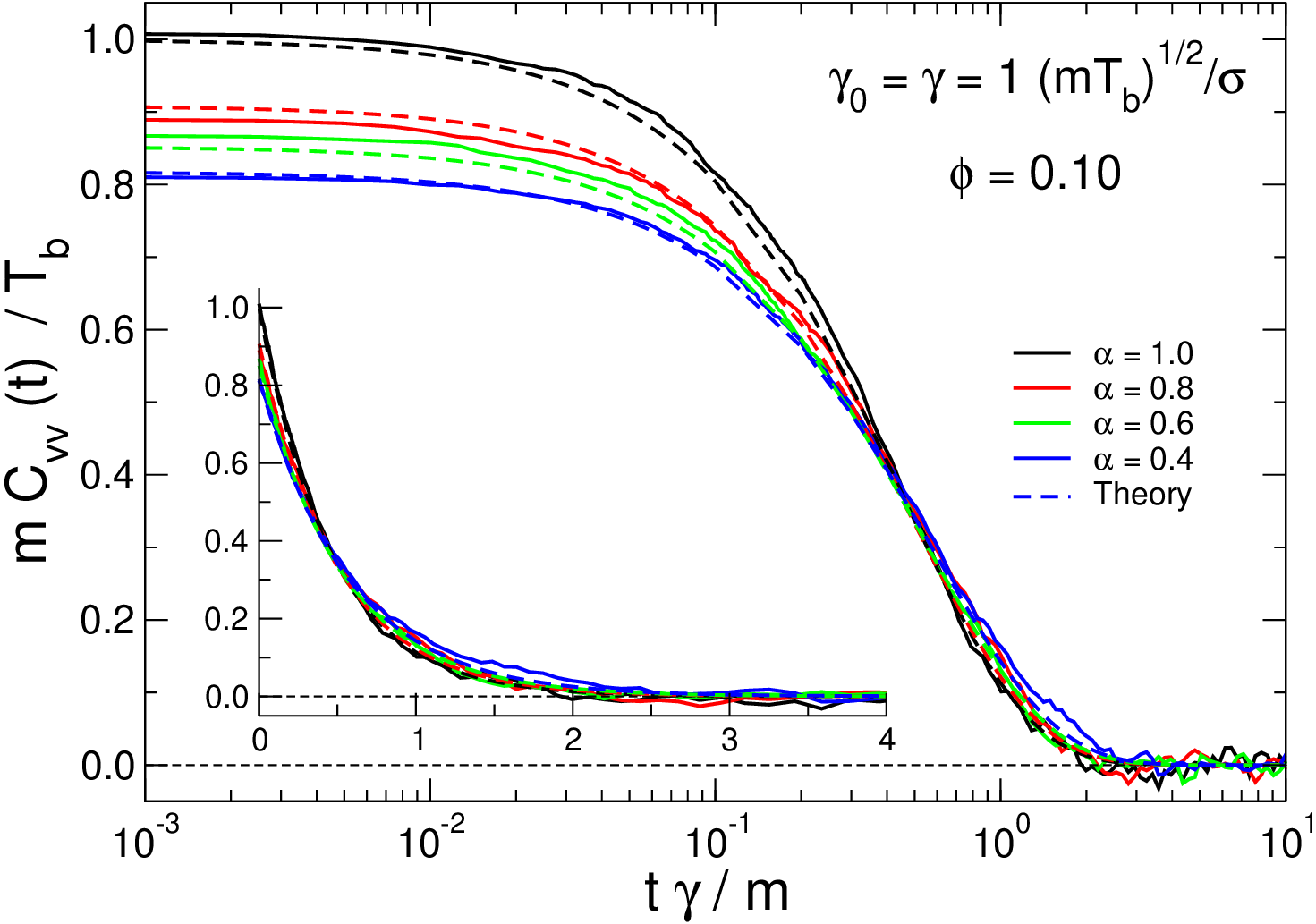}
    \includegraphics[width=\columnwidth]{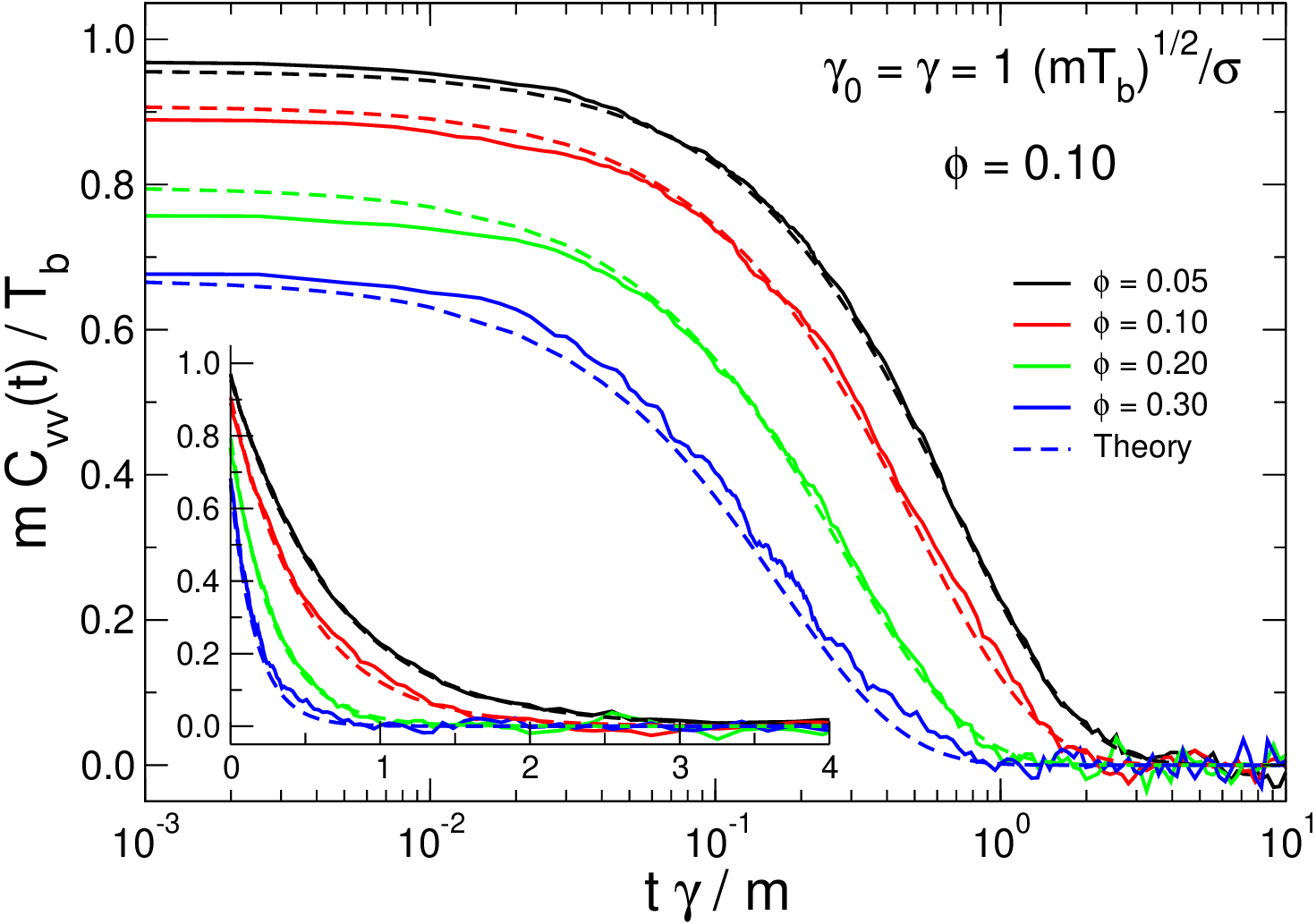}
    \caption{Velocity autocorrelation function. The upper panel shows it for different values of the restitution coefficient (volume fraction $\phi=0.10$ in all cases). The lower panel depicts it for different volume fractions (restitution coefficient, $\alpha=0.8$). MD simulations are represented by solid lines and theory by dashed lines.}
    \label{vvcorr}
\end{figure}

In order to study the dynamics of the system, Fig. \ref{vvcorr} shows the velocity autocorrelation functions, $C_{vv}(t)$ for different values of the restitution coefficient and volume fractions. The upper panel of Fig.\ \ref{vvcorr} shows the dependence of the velocity autocorrelation function $C_{vv}(t)$ on the restitution coefficient $\alpha$. A crossing of the different curves is observed, resulting from the interplay between the initial amplitude and the decay rate. Lower values of $\alpha$ reduce the initial correlations due to stronger dissipation but lead to a slower decay as the collision frequency decreases over time, causing the curves to intersect at intermediate times. Furthermore, the VACF are more stretched for smaller $\alpha$, resulting in a larger diffusion coefficient. This behavior can be interpreted within a random-walk framework \cite{GABG23}, where it is attributed to the reduction of the effective mean free path with increasing $\alpha$, which in turn diminishes the persistence of the intruders' trajectories.  

The lower panel, on the other hand, shows the effect of the volume fraction for an inelastic case. Increasing $\phi$ not only reduces the amplitude of the VACF but also accelerates its decay, as expected from the enhanced collision frequency at higher densities. This behavior reflects the increasing role of collisions at higher densities: the excluded-volume constraints and more frequent interactions enhance momentum transfer, leading to a faster loss of velocity correlations. 

Additionally, the comparison between theory and simulations shows very good overall agreement for the VACF, although small differences are visible, especially at higher volume fractions. Since the diffusion coefficient $D$ is obtained from the time integral of the VACF, these deviations result in quantitative discrepancies between the theoretical and simulated values of $D$, at least within the first Sonine approximation, although the qualitative trends remain consistent.

\begin{figure}
    \includegraphics[width=\columnwidth]{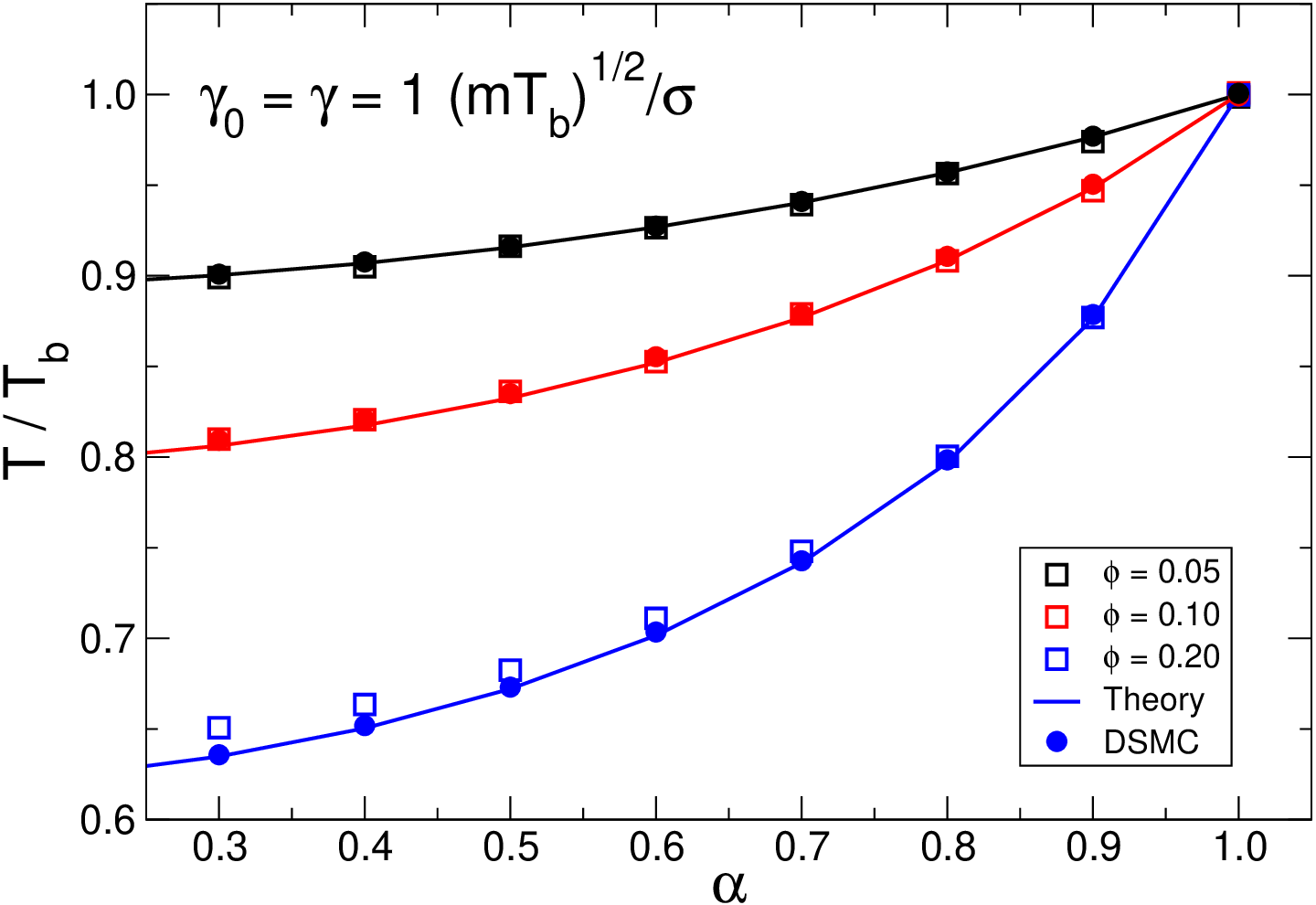}
    \caption{Temperature as a function of $\alpha$ for different volume fractions $\phi$, as labeled. Open symbols, lines, and closed symbols represent the simulation data, theory, and DSMC calculations, respectively. The error bars in the MD simulation results are smaller than the symbol size.}
    \label{ekin-alpha}
\end{figure}

The VACF encodes the most relevant variables describing the dynamics of the system, namely, the granular temperature and diffusion coefficient, as discussed previously. 
Figure \ref{ekin-alpha} shows the former as a function of $\al$, obtained from theory, MD, and DSMC simulations (in MD, the temperature is averaged over all particles in the system, as all of them are identical, to improve the statistics).The figure shows $T/T_\text{b}$ as a function of the restitution coefficient and volume fraction, displaying good overall agreement between theory and simulations, except for slight deviations at strong inelasticity and moderate densities. Therefore, the simulations confirm the presence of energy non-equipartition in granular suspensions.  
Here, the bath temperature $T_\text{b}$ sets the energy scale of the external thermal reservoir and provides the driving mechanism that compensates for collisional dissipation. As a result, the system does not relax toward thermal equilibrium ($T=T_\text{b}$), but instead reaches a nonequilibrium steady state in which the granular temperature $T$ is fixed by the balance between dissipation and energy injection. This leads to a stationary temperature ratio $T/T_\text{b}\neq 1$, which depends on the restitution coefficient and density, and quantifies the degree of non-equipartition between the grains and the background fluid.

\begin{figure}
    \includegraphics[width=\columnwidth]{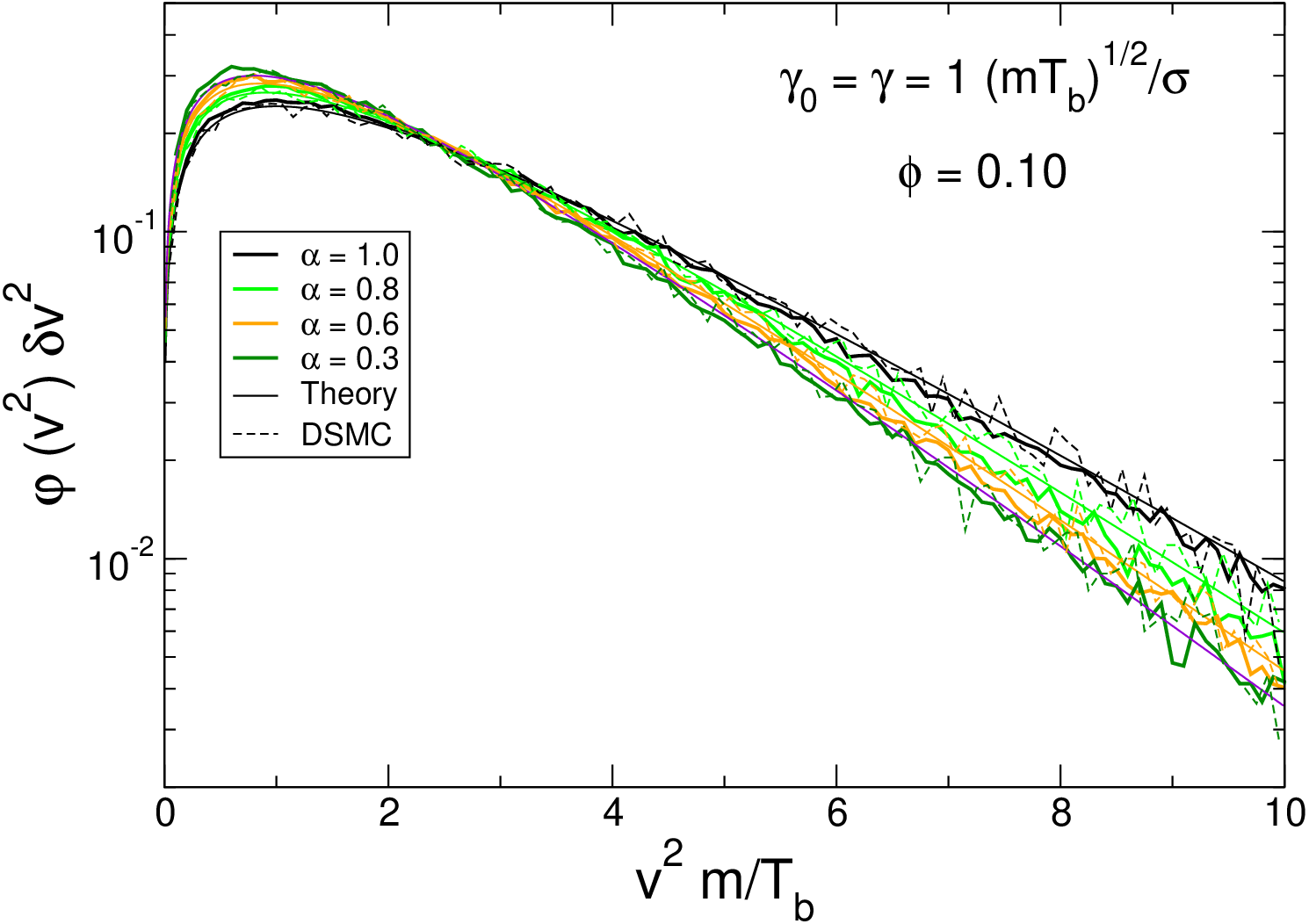}
    \caption{Distribution of the squared velocity for $\phi=0.1$ and different values of $\alpha$, as labeled. The thick solid lines represent the MD simulation results, dashed lines DSMC data, and thin solid lines stand for theory.} 
    \label{distr-T-alpha}
\end{figure}

To study the deviation from equipartition, the distribution of the squared velocity is shown in Fig.\ \ref{distr-T-alpha} for selected values of $\alpha$. The distribution shows a maximum for $v^2 \approx T/m$ and a quasi-exponential tail for large values. Upon decreasing $\alpha$, the peak increases and shifts to smaller values, while the tail decays faster. We compare the simulation results with the theoretical prediction of the function $\varphi(v^2)$, obtained from the Maxwell-Boltzmann velocity distribution after changing variables from the vectorial distribution $f(\mathbf{v})$ [Eq.\ \eqref{6.4}] to the scalar $v^2$.

\beq
\varphi(v^2) = 2\pi \left( \frac{m}{2\pi T} \right)^{3/2}v \exp\left( -\frac{m v^2}{2T} \right).
\eeq

The simulations results (thick lines) are perfectly matched by the theoretical predictions (thin lines). This reinforces the use of Maxwellian distributions and the fact that the cumulants are negligible, once again validating the results obtained in Refs.\ \cite{KG14,GGG21}, where it was shown that the thermalization imposed by the bath drives the system to behave with dynamics closer to equilibrium than in the case of HCS.
\vspace{1cm}
\begin{figure}
    \includegraphics[width=\columnwidth]{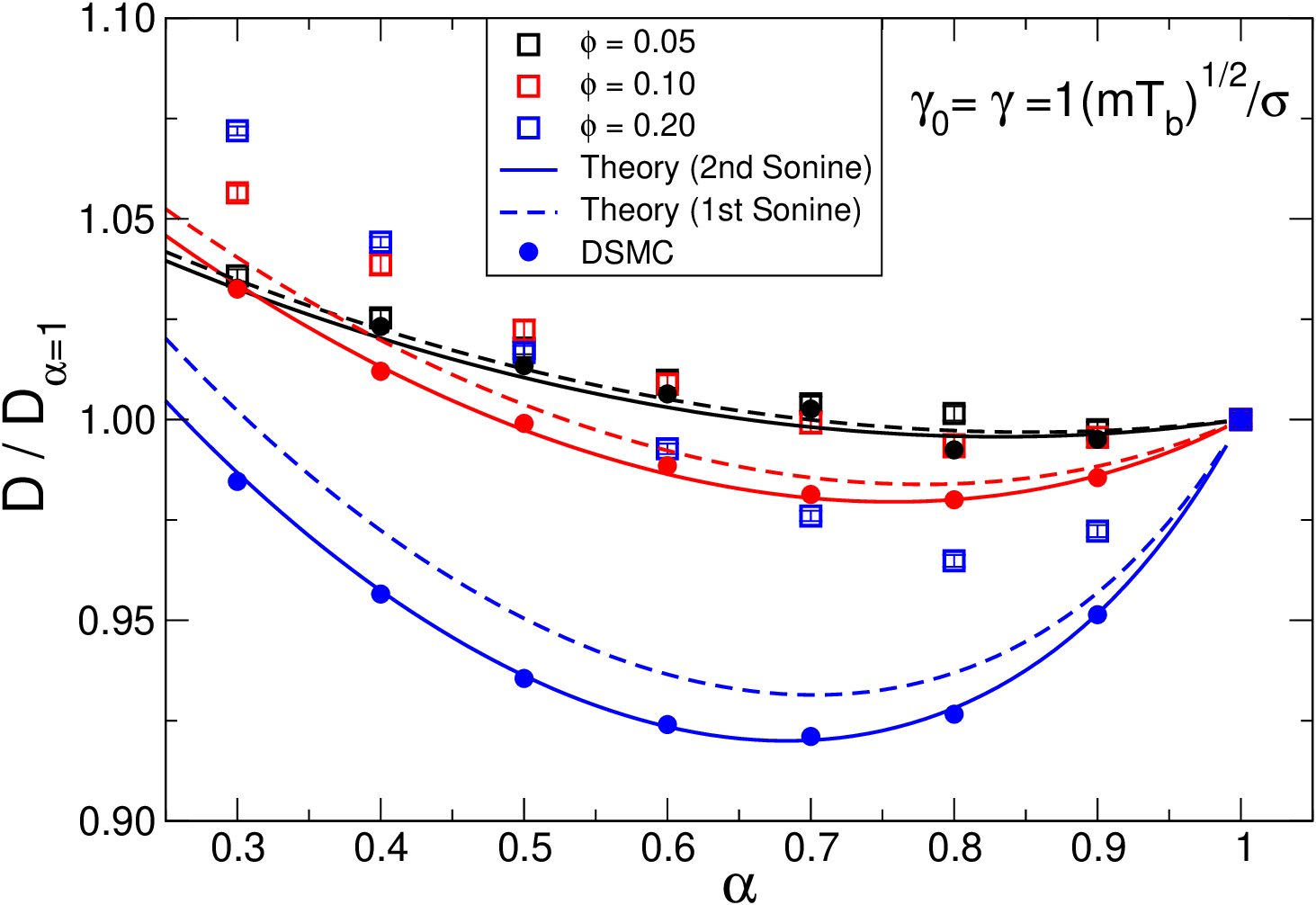}
    \caption{Diffusion coefficient as a function of $\alpha$ for different volume fractions $\phi$, as labeled. Open symbols, lines, and closed symbols represent the simulation data, theory, and DSMC calculations, respectively. Error bars in the simulation data correspond to the standard error of independent determinations of $D$.}
    \label{D-alpha}
\end{figure}

Focusing on the diffusion coefficient, it displays a non-trivial behavior, as shown in Fig.\ \ref{D-alpha}, where it is plotted for different volume fractions and normalized by the corresponding value for elastic collisions, and given in Table \ref{table-D}. For the lowest volume fraction considered, simulations and theory are in good agreement, both showing that the coefficient increases as $\alpha$ decreases. At higher densities, however, a minimum appears before the diffusion coefficient reaches the value corresponding to elastic collisions.
Comparison with the simulation results shows that the theory qualitatively reproduces the observed non-monotonic behavior. This trend arises from the competition between two opposing effects: the reduction in trajectory persistence as $\alpha$ increases, and the simultaneous rise in the collision rate, which hinders diffusion. 

\begin{widetext}
\begin{table*}
\centering
\begin{tabular}{ccccccccccc} \hline
$\alpha$ & \multicolumn{3}{c}{$\phi=0.05$} & \multicolumn{3}{c}{$\phi=0.10$} & \multicolumn{4}{c}{$\phi=0.20$} \\ \hline
   & $2^{nd}$ Sonine & DSMC & MD  & $2^{nd}$ Sonine & DSMC &  MD  & $2^{nd}$ Sonine & DSMC & MD  &  MD ($N=8000$) \\ \hline
1.0 & 0.6616 & 0.6626 & 0.6555(5) & 0.4610 & 0.4680 & 0.4588(6) & 0.2415 & 0.2447 & 0.2519(2) & 0.2513(2) \\ 
0.9 &  0.6592 & 0.6594 & 0.6529(6) & 0.4546 & 0.4612 & 0.4568(6) &  0.2298 & 0.2328 & 0.2449(4) & 0.2441(2) \\
0.8 & 0.6589  & 0.6576 & 0.6555(6) &  0.4518& 0.4586 & 0.4558(9) & 0.2241 & 0.2267 & 0.2431(4) & 0.2431(1) \\
0.7 & 0.6604 & 0.6644 & 0.6571(5) & 0.4520  & 0.4592 & 0.4584(8) & 0.2222 & 0.2253 & 0.2459(3) & 0.2443(3) \\
0.6 & 0.6636 & 0.6669 & 0.6610(6) & 0.4548 & 0.4626 & 0.4628(7) & 0.2230 & 0.2261 & 0.2501(3) &  0.2489(2) \\
0.5 & 0.6685 & 0.6715  & 0.6662(4) & 0.4598 & 0.4675 & 0.4690(8) & 0.2261 & 0.2289 & 0.2561(5) & 0.2548(2) \\
0.4 & 0.6750 & 0.6780& 0.6712(6) & 0.4671 & 0.4736 & 0.4765(8) & 0.2312 & 0.2340 & 0.2631(3) &  0.2624(2)\\
0.3 & 0.6831 & 0.6843 & 0.6780(4) & 0.4766 & 0.4832 & 0.4847(3) & 0.2383 & 0.2408 & 0.2700(2) & 0.2743(7) \\  \hline
\end{tabular}
\caption{Second Sonine, DSMC, and MD values of the diffusion coefficient $m\gamma D/T_\text{b}$ for different values of the coefficient of restitution $\alpha$ and volume fraction $\phi = 0.05, 0.10,$ and $0.20$. DSMC results are shown without standard errors since they are smaller than the significant digits reported for the MD data. MD results correspond to simulations with $N = 1000$ particles, except for the last column at $\phi = 0.20$, where results for $N = 8000$ are explicitly indicated. Simulation values and standard errors are rounded to the last significant digit.}
\label{table-D}
\end{table*}
\end{widetext}

Interestingly, Fig.\ \ref{D-alpha} shows that the minimum coefficient obtained from DSMC (or Enskog kinetic theory) systematically shifts toward higher values of $\alpha$ compared to the simulations. This systematic shift suggests that Enskog's description slightly underestimates the persistence of the intruder's trajectories. Consequently, higher collision frequencies are required in the theoretical framework to recover the mean square displacement, leading to the rightward shift of the minimum. Quantitatively, some discrepancies are observed between the MD and theoretical results at higher densities and/or strong inelasticities, probably due to a breakdown of molecular chaos. In particular, the Chapman--Enskog approach may lose accuracy in dense or strongly correlated regimes, although the velocity autocorrelation function itself shows very good agreement. It should also be noted that the normalization used
%$D$ is scaled by its value for elastic collisions, in order to present results for different volume fractions using the same scale. This normalization, however, 
amplifies the apparent quantitative differences between the curves, which nevertheless remain below 8\% for the highest volume fraction. Table \ref{table-D} shows the values of the diffusion coefficient to allow for a direct comparison, as well as the results from MD simulations with $N=8000$ particles, to confirm that the observed differences are not caused by the finite size of the simulated system.
Finally, it is noticeable that the second Sonine approximation shows a markedly improved agreement with the DSMC results, confirming that it provides a more accurate representation within the Enskog framework. For this reason, the theoretical results presented hereafter are based on the second Sonine approximation, which provides a more reliable description.

In order to test the molecular-chaos hypothesis, the spatial correlation of particle displacements is measured as proposed in Ref. \cite{Reza2020} using the following two-particle function:

\beq
\Delta(r) = \frac{\langle \mathbf{\delta_1} \cdot \mathbf{\delta_2} \rangle}{\delta_1^2},
\eeq

\begin{figure}
    \includegraphics[width=\columnwidth]{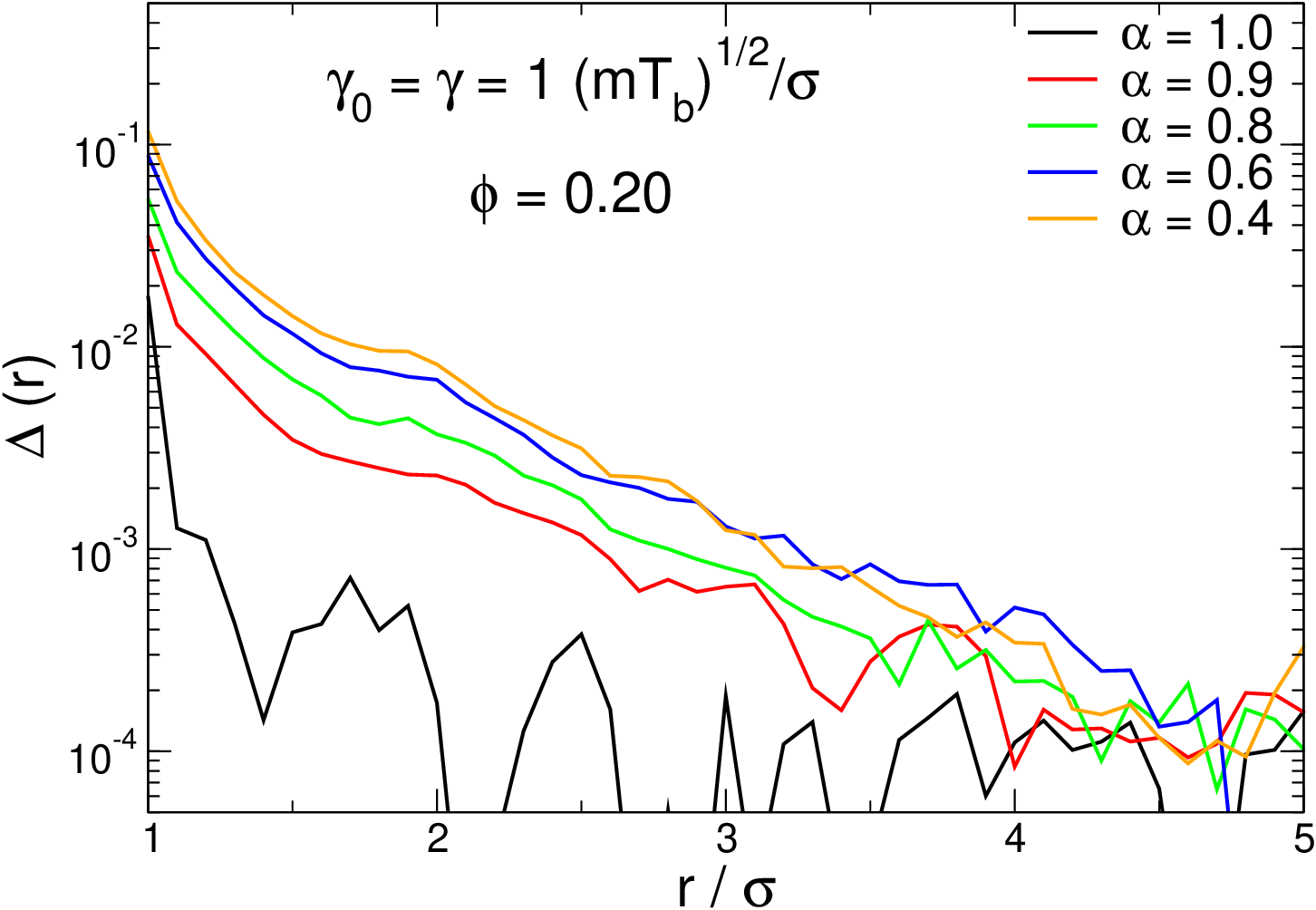}
    \caption{Correlated motion obtained from MD simulations for $\phi=0.2$ and different values of $\alpha$, as labeled}
    \label{d12}
\end{figure}

\noindent where $\mathbf{\delta_1}$ and $\mathbf{\delta_2}$ refer to the displacement of particle $1$ and $2$, respectively, over a time interval of five time steps in our case, and $r=\left| \mathbf{r}_2 - \mathbf{r}_1 \right|$ is the separation between the particles. The results are shown in Figure \ref{d12} for $\phi=0.20$ and different values of $\alpha$. In the elastic case, the function decays very fast, showing the absence of correlated motions. In contrast, upon decreasing $\alpha$, the correlations become more relevant and extend to long distances. This behaviour differs from the molecular-chaos, and therefore deviations from results based on the Chapman--Enskog approach are expected for small $\alpha$.

To further investigate whether correlated motion is the origin of the discrepancy  between the diffusion coefficients from MD simulations and the theoretical predictions,  we varied the friction coefficient $\gamma$, which controls the strength of the Brownian  forcing and thus the randomization of particle motion. We observe that the correlation  function $\Delta(r)$ decays more rapidly for larger $\gamma$, and the diffusion coefficient 
$D$ from simulations approaches the theoretical values, particularly at low $\alpha$. This indicates that the discrepancies may arise from correlated particle motion, which is absent in the Enskog kinetic theory.

%It is also possible that these deviations are amplified by representing $D$ relative to its elastic value, which accentuates the discrepancies. We have verified this point and indeed the deviations decrease when plotting $D\gamma/T_\text{b}$. However, this representation obscures the influence of $\alpha$, which is why we decided to keep Fig.\ \ref{ekin-D-alpha} in its current form.}

\subsection{Effect of the intruder mass}

\begin{figure}
    \includegraphics[width=\columnwidth]{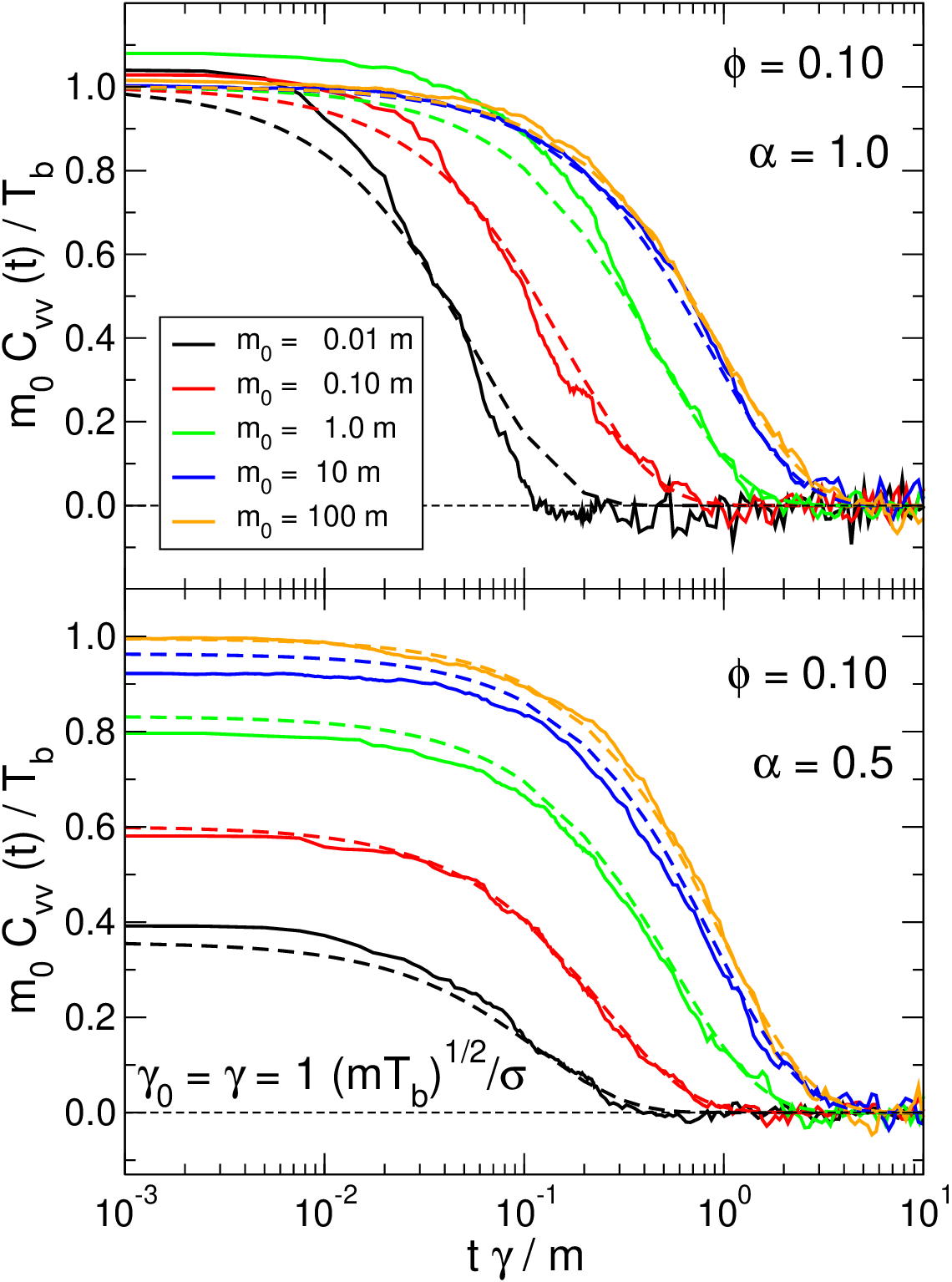}
    \caption{Intruder velocity autocorrelation function for $\phi=0.1$ and different values of the intruder mass, $m_0$, as labeled. Two values of $\alpha$ have been considered: $\alpha=1.0$ in the upper panel, and $\alpha=0.5$ in the lower one. The solid lines represent the MD results and dashed lines stand for theory.}
    \label{vvcorr-mt}
\end{figure}

Next, we perform simulations of the system with an intruder mechanically different from the granular particles. Figure \ref{vvcorr-mt} shows the VACF for intruders of different masses, and for two values of $\al$. In the elastic case ($\alpha=1$), the intruder mass only modifies the time scale of the decay: all curves start with the same amplitude and differ mainly in how fast correlations vanish, so that increasing $m_0$ results in a slower relaxation without modifying the functional shape of the VACF. Conversely, for $\alpha=0.5$ both the amplitude and the slope are affected, indicating that the intruder does not reach equipartition with the bath and that its kinetic energy depends strongly on $m_0$. This distinction implies that, while in the elastic system intruder dynamics are governed solely by inertia, inelasticity introduces an additional dissipative mechanism that accelerates decorrelation for light intruders and enhances memory retention for heavy ones. We again observe good agreement between the MD results and the theoretical predictions, although small discrepancies appear for low mass ratios $m_0/m$, reflecting minor differences in the granular temperature and tracer diffusion coefficient (first Sonine approximation) when the intruder is lighter.

%in the elastic case (upper panel) the effect of $m_0$ is to increase the time scale of the decay of the VACF until it saturates for large masses, for $\alpha=0.5$ its amplitude is also affected, implying a change in the average kinetic energy of the tracer.

\begin{figure}
    \includegraphics[width=\columnwidth]{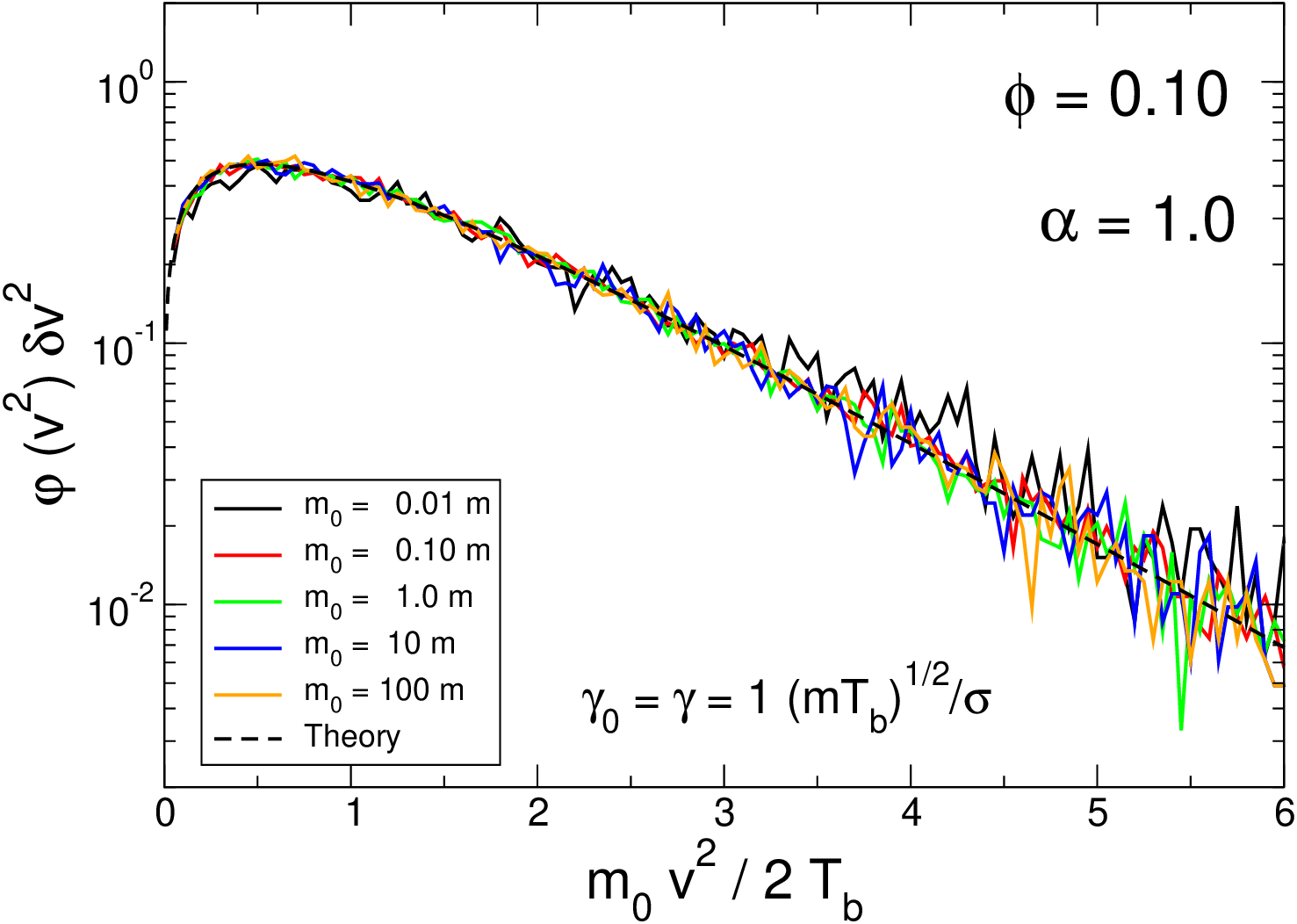}
    \includegraphics[width=\columnwidth]{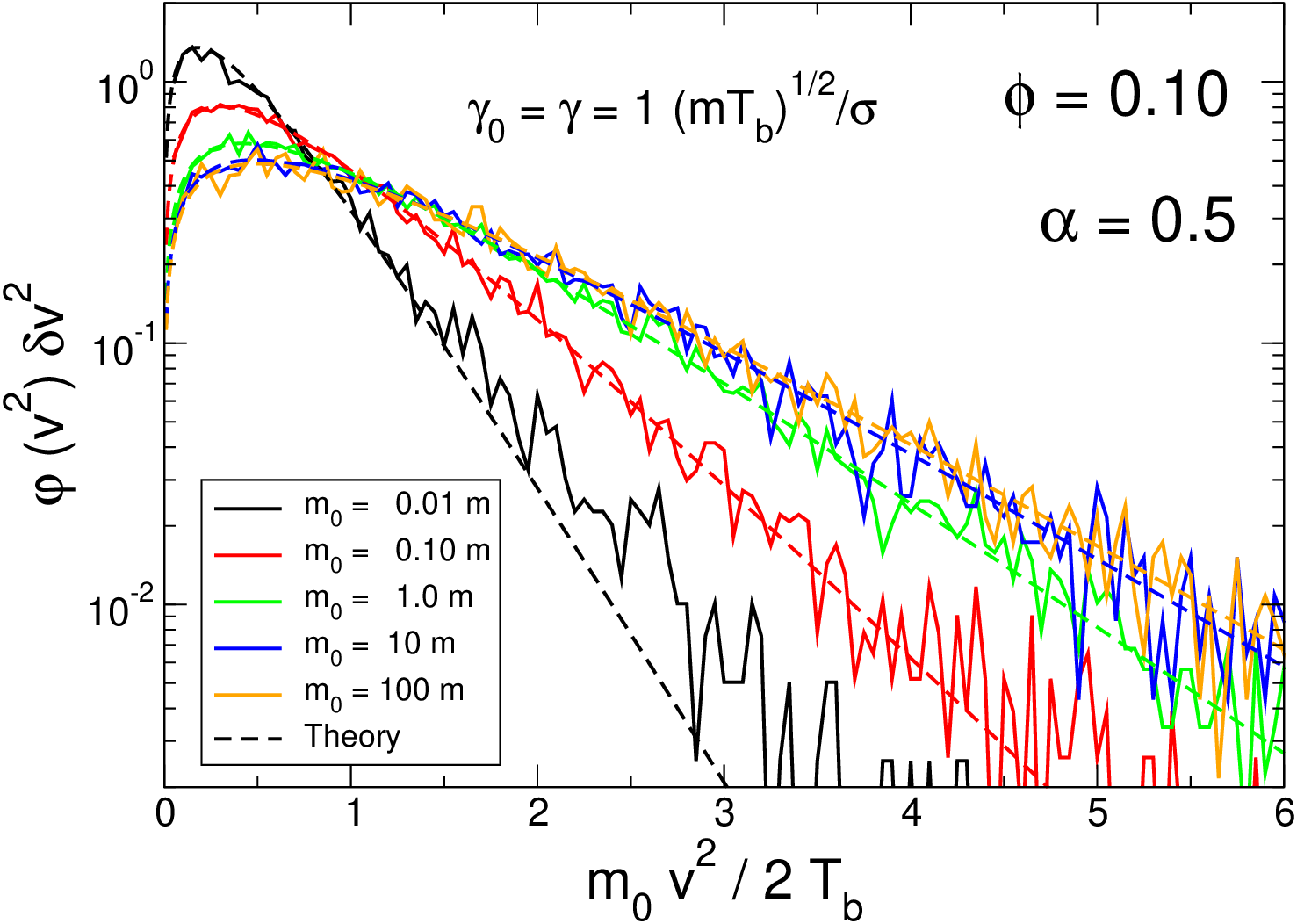}
    \caption{Distribution of the intruder kinetic energy for $\phi=0.1$ and different values of the intruder mass, as labeled. The upper panel corresponds to the $\al=1$ and the lower one to $\al=0.5$. The solid and dashed lines represent the distributions from MD simulations and theory, respectively. }
    \label{ekin-mt}
\end{figure}

The distribution of kinetic energy $\varphi(v^2)$ is shown in Fig.\ \ref{ekin-mt} for the same values of the intruder's mass, for the elastic case and $\al=0.5$ (upper and lower panels, respectively). As expected, in the elastic case, the distribution is stable upon changing the mass ratio. However, for $\alpha=0.5$, as $m_0$ increases the peak of the distribution shifts to larger values and the high-energy tail decays more slowly, tending to saturate for $m_0>m$. This peak shift reflects the increase of the intruder’s average kinetic energy with mass and the breakdown of equipartition, while the slower decay of the tail reveals a higher probability of energetic intruder particles. This behavior shows that heavier intruders, due to their inertia, preserve their trajectories more effectively, thereby enhancing diffusion. Moreover, the theory successfully reproduces both datasets across the entire range of intruder masses studied, particularly in the inelastic case. Although not shown, DSMC results also exhibit good agreement with both MD and theoretical predictions. 

\begin{figure}
    \includegraphics[width=\columnwidth]{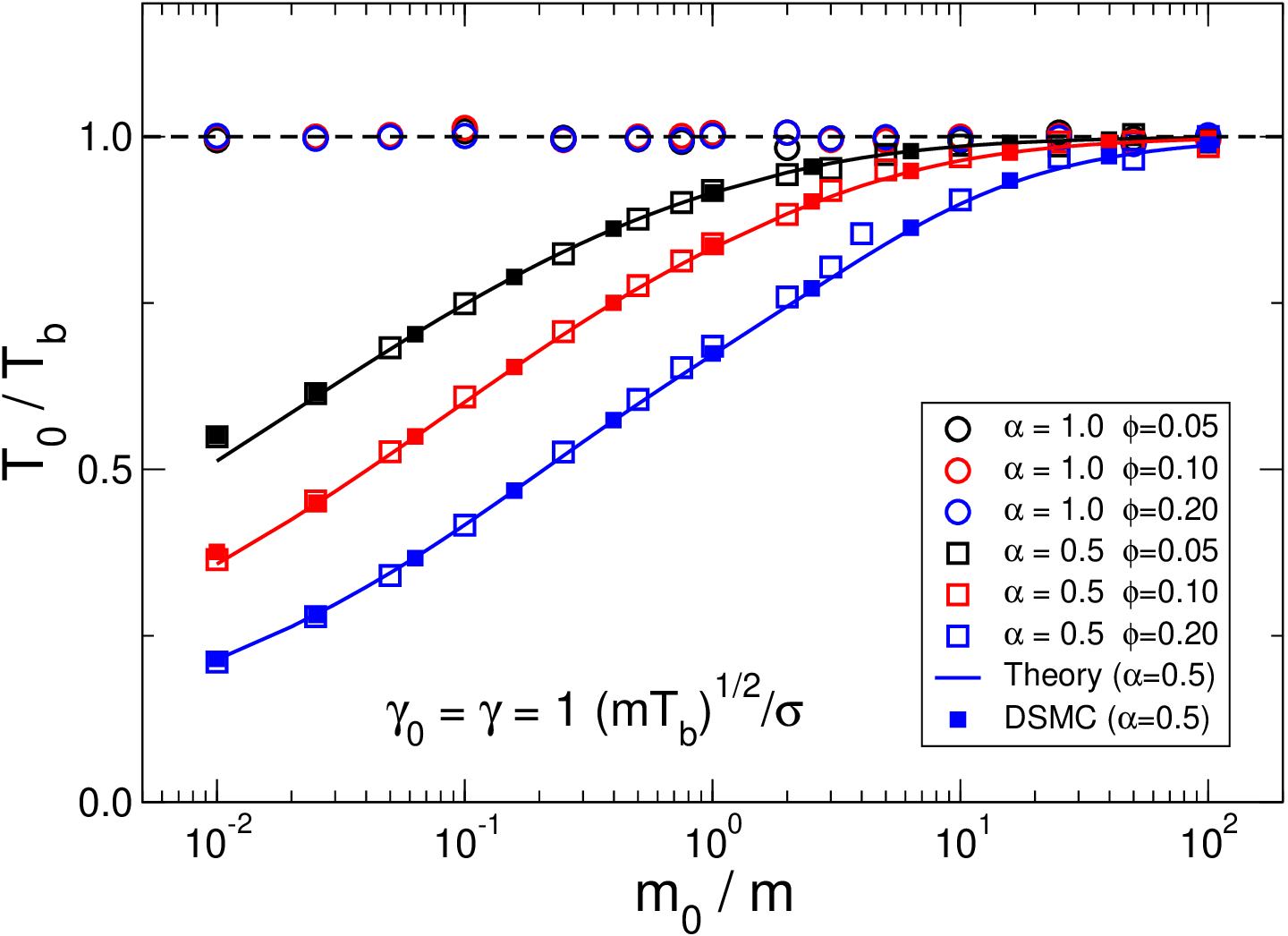}
    \includegraphics[width=\columnwidth]{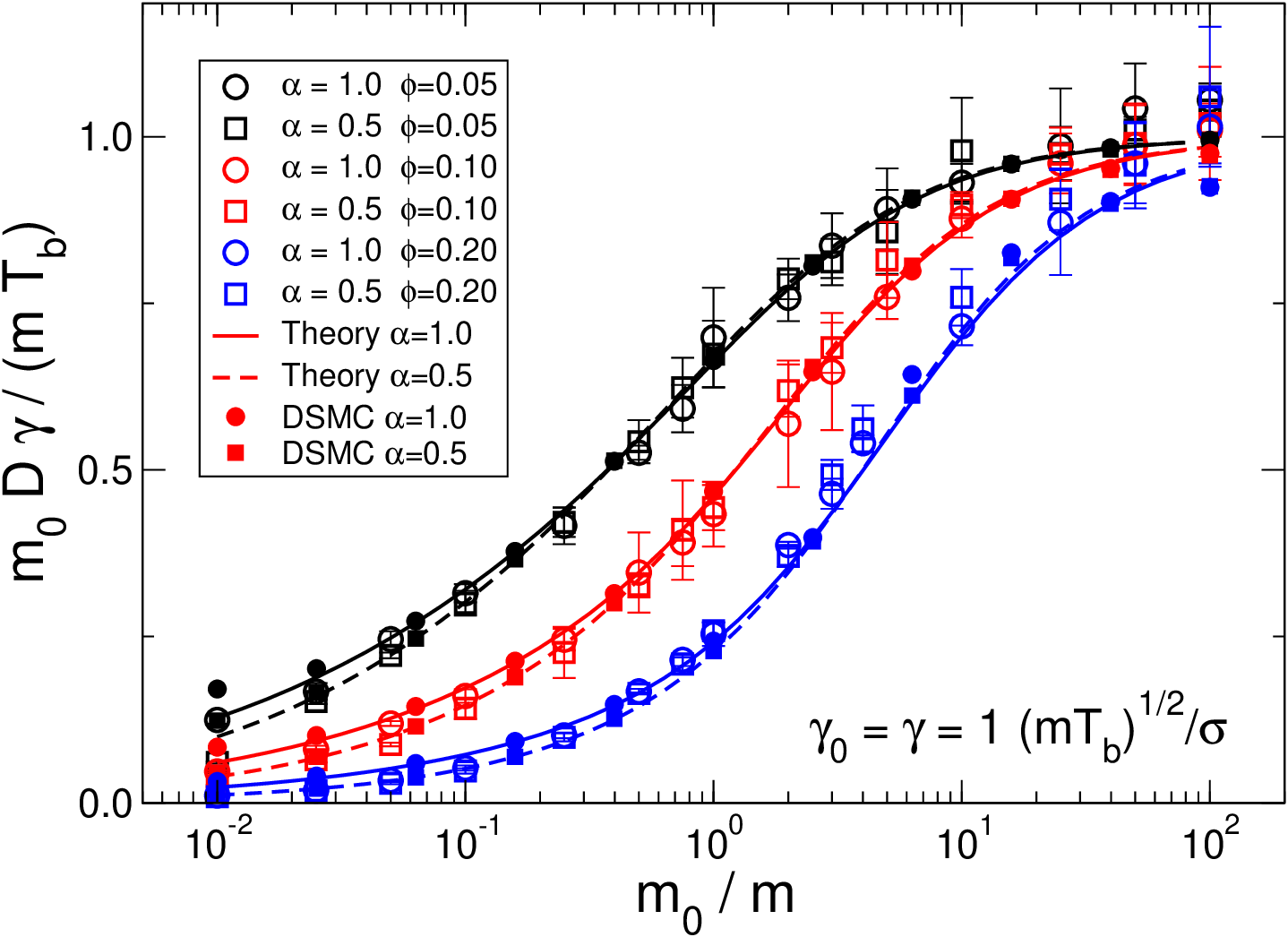}
    \caption{Granular temperature of the intruder (upper panel) and tracer diffusion coefficient (lower panel) as a function of the intruder mass, for different densities, as labeled. Results are shown for two restitution coefficients, $\alpha = 0.5$ and $\alpha = 1$. Open symbols, lines, and closed symbols represent the simulation data, the second Sonine approximation, and DSMC calculations, respectively. As in Fig. \ref{D-alpha}, the error bars in the temperature are smaller than the symbol size and in $D$ represent the standard error of the data for different independent evaluations.}
    \label{Dtracer-mt}
\end{figure}

The average intruder temperature is shown in the upper panel of Fig. \ref{Dtracer-mt}, where the equipartition theorem is fulfilled for $\al=1$, whereas deviations are observed for light intruders, increasing with the bath density. Again, as expected from the comparison with the theory in the previous figure, the model reproduces the simulation results perfectly. It is important to remark that the agreement for $T_0/T_b$ found here between kinetic theory and MD simulations is much better than the one reported previously in the case of the HCS (see figs. 2 and 3 of Ref.\ \cite{DHGD02})

The lower panel of Fig. \ref{Dtracer-mt} shows second Sonine approximation for the tracer diffusion coefficient $D$ with mass $m_0$ and the same diameter as the bath particles, with simulations and DSMC results. (Note that the diffusion coefficient is scaled with the mass ratio). For all volume fractions considered, $D\,m_0/m$ grows from values close to zero to one, with the crossover at $m_0 \approx m$, with a negligible effect from the restitution coefficient (identical for all particles). The comparison with the theory shows excellent agreement for all cases, as well as with the DSMC calculations. In Ref. \cite{GABG23}, it was shown that the second Sonine approximation significantly improves upon the first one for light intruders when compared against DSMC data. Here, we find that the second approximation provides optimal predictions across the entire mass-ratio range. For this reason, the first Sonine approximation is not represented in the figure. For massive intruders, these results indicate that the slow diffusion is accounted for with the mass ratio, recovering the free diffusion limit for all densities. This mechanism is correctly predicted by the theory.

\section{Conclusions}

A model for an intruder (or tracer) in a granular gas immersed in an interstitial solvent has been tested in this work using Langevin dynamics and direct simulation Monte Carlo (DSMC). The model accounts for inelastic collisions and a mixture of particles with different mechanical properties, with one species considered in the tracer limit. It is based on the Enskog equation, assuming that the interstitial gas is dilute enough so that the Enskog collision operator remains the same as for a dry granular gas. In this regime, the effect of the solvent on the particles is modeled as an external force composed of two terms: a drag force proportional to particle velocity and a stochastic Langevin-like term.

Our objective is to employ the Chapman--Enskog expression for the tracer diffusion coefficient, derived in Ref.~\cite{GABG23} up to the second Sonine approximation, to test the predictions of Enskog kinetic theory against molecular dynamics simulations. The theoretical results rely on several standard approximations, including low-to-moderate densities, molecular chaos, and truncation of the Sonine expansion, and the comparisons presented here provide strong support for their validity. The simulations directly probe these assumptions by integrating the same equation under a variety of conditions, confirming their applicability for the suspension model and offering a solid foundation for future extensions. Notably, the stochastic forces in the Langevin equation introduce an additional decorrelation mechanism for velocity correlations beyond collisions, further enhancing the predictive power of kinetic theory.
 %Furthermore, the fluctuation--dissipation theorem establishes a thermodynamic connection to the granular temperature. 

The effect of the drag coefficient was first examined to identify a regime where comparison with theory is meaningful. We found that for low-to-intermediate values of the drag parameter, the theory accurately predicts simulation results. For the remainder of the study, we set $\gamma\sigma/(mT_\text{b}) = 1$, where the solvent’s influence is comparable to that of particle collisions.

Subsequently, the intruder velocity autocorrelation function was analyzed under various conditions, including changes in the restitution coefficient, volume fraction, and intruder mass, and compared with theoretical predictions. The granular temperature, the tracer diffusion coefficient, and the velocity distribution were also analyzed. Overall, the agreement between theory and simulations is good, with
only quantitative discrepancies in the diffusion coefficient as a function of inelasticity. The discrepancies depend on the order of the Sonine expansion: the second Sonine approximation closely matches the DSMC results, indicating that the theoretical assumptions underlying the Enskog equation are valid at this level, and thus constitutes the most accurate representation. These results strongly support the use of kinetic theory to describe the dynamic properties of a intruder in a granular suspension, confirming its validity across a wide range of densities, intruder masses, and degrees of inelasticity.

A natural extension of this study is the analysis of granular suspensions under shear flow, for which the isotropic and homogeneous steady state reported here serves as a fundamental reference. The excellent agreement found between kinetic theory and both MD and DSMC simulations motivates the exploration of anisotropic transport in far-from-equilibrium states. In this context, recent work on inertial suspensions under simple shear has characterized the emergence of anisotropic diffusion tensors within the Langevin–Enskog framework \cite{THG23}. While the present study focuses on isotropic conditions, it provides the necessary baseline to assess such anisotropic extensions. Furthermore, kinetic theories for tracer diffusion in sheared granular gases \cite{MG26} offer complementary insights into nonlinear transport under strong shear. Validating these theoretical frameworks for sheared suspensions against MD simulations constitutes a clear direction for future research.

\begin{acknowledgments}
The authors gratefully acknowledge the helpful comments and fruitful discussions with Enrique Abad, Santos B. Yuste, and Vicente Garz\'o. Their insights have greatly contributed to improving this work. A.M.P. also acknowledges financial support from the Spanish Ministerio de Ciencia, Innovación y Universidades, Spain, through project No.
PID2021-127836NBI00 (funded by MCIN/AEI/10.13039/501100011033/FEDER ‘‘A way to make Europe’’).
\end{acknowledgments}

\appendix
\section{Collision frequencies of $D$}
\label{appA}

To obtain the explicit dependence of $D[1]$ and $D[2]$ on the system parameters, it is necessary to know the collision frequencies $\nu_a$, $\nu_b$, $\nu_c$, and $\nu_d$. These quantities have been previously calculated under the assumption that the velocity distribution function $f$ is approximated by the Maxwellian form \eqref{6.4}, as shown in Refs.\ \cite{GM07,GHD07,GV09}. For completeness, we reproduce the corresponding expressions below. A detailed derivation and recent discussion can also be found in \cite{GABG23}.

\begin{widetext}
\begin{equation}
\label{a1}
\nu_{a}=\frac{2\sqrt{2}\pi^{(d-1)/2}}{d\Gamma\left(\frac{d}{2}\right)}
\chi_0\mu (1+\alpha_0) \left(1+\beta\right)^{1/2}\left(\frac{T_0}{m_0}\right)^{1/2}n\overline{\sigma}^{d-1},
\end{equation}
\begin{equation}
\label{a2}
\nu_{b}=\frac{\sqrt{2}\pi^{(d-1)/2}}{d\Gamma\left(\frac{d}{2}\right)}
\chi_0\mu (1+\alpha_0) (1+\beta)^{-1/2}\left(\frac{T_0}{m_0}\right)^{1/2}n\overline{\sigma}^{d-1},
\end{equation}
\begin{equation}
\label{a3}
\nu_{c}=\frac{2\sqrt{2}\pi^{(d-1)/2}}{d(d+2)\Gamma\left(\frac{d}{2}\right)}
\chi_0\mu (1+\alpha_0) (1+\beta)^{-1/2}\left(\frac{T_0}{m_0}\right)^{1/2}A_c n\overline{\sigma}^{d-1},
\end{equation}
\begin{equation}
\label{a4}
\nu_{d}=\frac{\sqrt{2}\pi^{(d-1)/2}}{d(d+2)\Gamma\left(\frac{d}{2}\right)}
\chi_0\mu (1+\alpha_0)\frac{\beta}{(1+\beta)^{3/2}}\left(\frac{T_0}{m_0}\right)^{1/2}
\left[A_d-(d+2)\frac{1+\beta}{\beta} A_c\right]n\overline{\sigma}^{d-1},
\end{equation}
where
\begin{eqnarray}
\label{a5}
A_c&=& (d+2)(1+2\lambda)+\mu(1+\beta)\Big\{(d+2)(1-\alpha_0)
-[(11+d)\alpha_0-5d-7]\lambda\beta^{-1}\Big\}+3(d+3)\lambda^2\beta^{-1}\nonumber\\
& &+2\mu^2\left(2\alpha_0^{2}-\frac{d+3}{2}\alpha
_{12}+d+1\right)\beta^{-1}(1+\beta)^2- (d+2)\beta^{-1}(1+\beta),
\end{eqnarray}
\begin{eqnarray}
\label{a6}
A_d&=&2\mu^2\left(\frac{1+\beta}{\beta}\right)^{2}
\left(2\alpha_0^{2}-\frac{d+3}{2}\alpha_0+d+1\right)
\big[d+5+(d+2)\beta\big]-\mu(1+\beta) \Big\{\lambda\beta^{-2}[(d+5)+(d+2)\beta]
\nonumber\\
& & \times
[(11+d)\alpha_0
-5d-7]-\beta^{-1}[20+d(15-7\alpha_0)+d^2(1-\alpha_0)-28\alpha_0] -(d+2)^2(1-\alpha_0)\Big\}
\nonumber\\
& & +3(d+3)\lambda^2\beta^{-2}[d+5+(d+2)\beta]+ 2\lambda\beta^{-1}[24+11d+d^2+(d+2)^2\beta]
\nonumber\\
& & +(d+2)\beta^{-1} [d+3+(d+8)\beta]-(d+2)(1+\beta)\beta^{-2}
[d+3+(d+2)\beta].\nonumber\\
\end{eqnarray}
Here, $\lambda=(\mu_0/T_0)\left(T_0-T\right)$.
\end{widetext}


%\bibliography{Brown}% Produces the bibliography via BibTeX.

\end{document}